\documentclass[twoside,slac_one]{revtex4}
\usepackage{graphicx}
\usepackage{fancyhdr}
\usepackage{atlasphysics}
\usepackage{amsmath} 
\usepackage{bm}
\usepackage{amsxtra}
\usepackage{amssymb}
\usepackage{amsthm}
\usepackage{latexsym}
\usepackage{lscape}
\usepackage{lineno}

\begin{document}

\title{Recent Heavy Ion Results with the ATLAS Detector at the LHC}

%

\author{P. Steinberg}
\affiliation{Department of Physics, Brookhaven National Laboratory, Upton, NY, USA}

\begin{abstract}
Results are presented from the ATLAS collaboration from the 2010 LHC heavy ion run,
during which nearly 10 inverse microbarns of luminosity were delivered.  
Soft physics results include charged particle multiplicities and 
collective flow.
The charged particle multiplicity, which tracks initial state entropy production,
increases by more than a factor of two relative to the top RHIC energy, with a 
centrality dependence very similar to that previously measured at RHIC.
Measurements of elliptic flow out to large transverse
momentum also show similar results to what was measured at RHIC.
%
Extensions of these measurements to higher harmonics have also been made,
and can be used to explain structures in the two-particle correlation 
functions that had long been attributed to jet-medium interactions.
New hard probe measurements include single muons, jets and high $p_T$ hadrons.
Single muons at high momentum are used to extract the yield of $W^{\pm}$ bosons
and are found to be consistent within statistical uncertainties 
with binary collision scaling.
Conversely, jets are found to be suppressed in central events by a factor of
two relative to peripheral events, 
with no significant dependence on the jet energy.  Fragmentation functions
are also found to be the same in central and peripheral events.
Updated asymmetry results confirm previous measurements with increased statistics, and 
multiple jet cone sizes are presented.
Finally, charged hadron spectra 
have been measured out to 30 GeV, and their centrality
dependence relative to peripheral events is similar to that found for jets.

\end{abstract}

\maketitle


\thispagestyle{fancy}


\section{Heavy ions with the ATLAS detector}

\begin{figure}[b]
\begin{center}
\includegraphics[width=19pc]{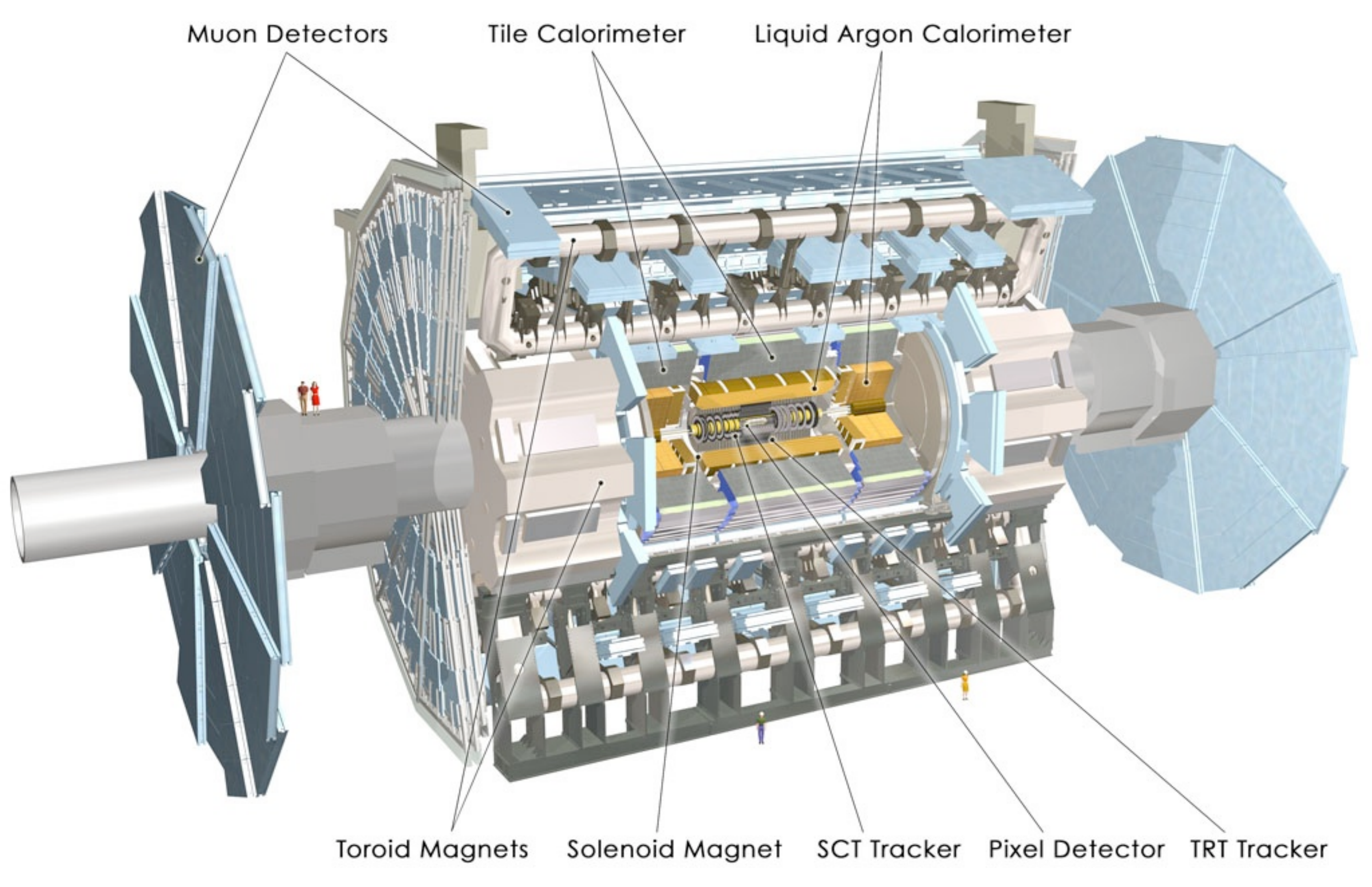}
\includegraphics[width=16pc]{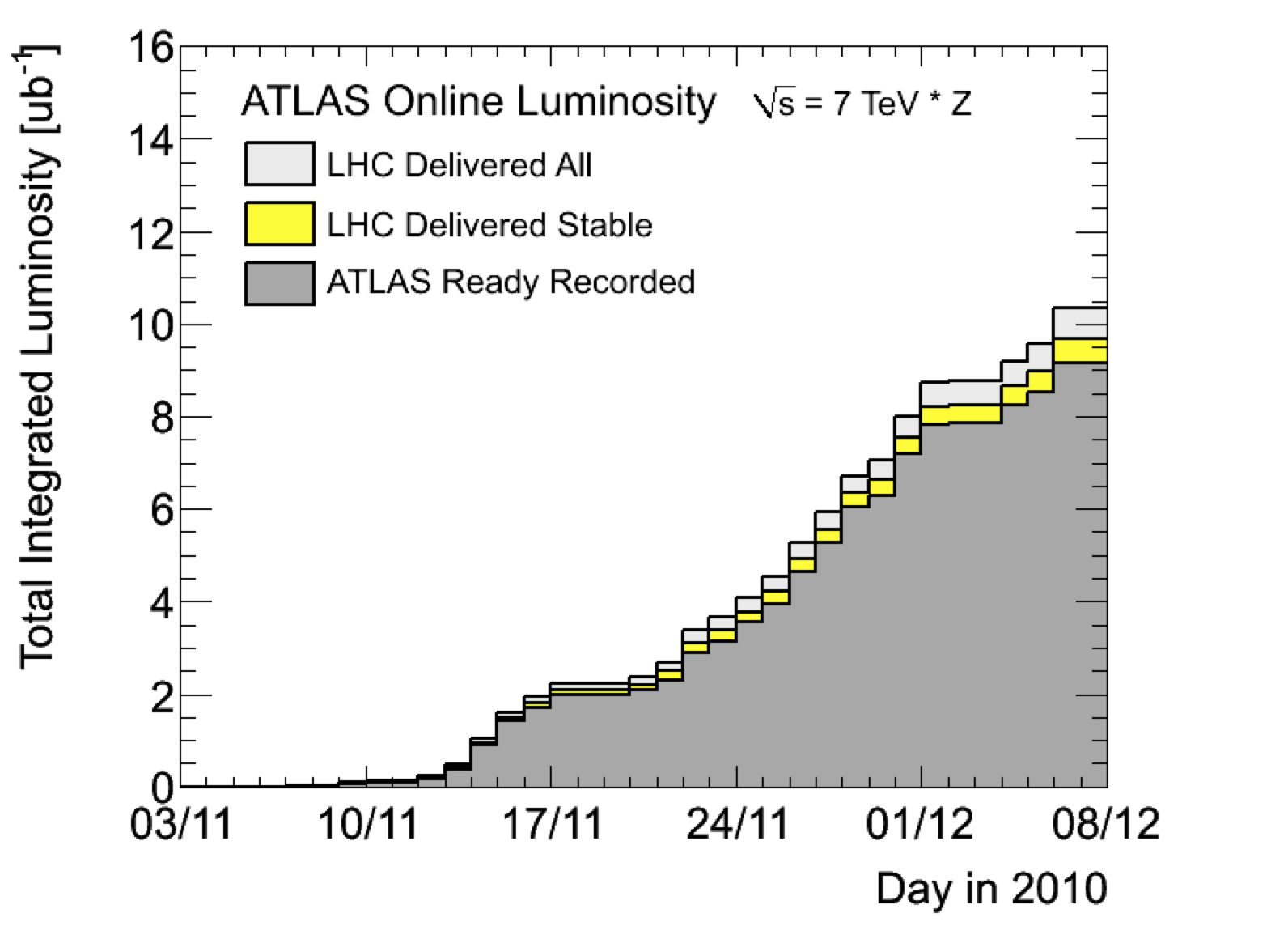}
\caption{\label{atlas}
(left) Schematic diagram of the ATLAS detector, showing the three main subsystems: the\
 inner detector ($|\eta|<2.5$),
the calorimeter ($|\eta|<4.9$) and the muon spectrometer ($|\eta|<2.7$).
(right) Integrated luminosity taken by ATLAS in the 2010 heavy ion run vs. time.
}
\end{center}
\end{figure}

The ATLAS detector at the LHC~\cite{Aad:2008zzm}, shown in Fig.~\ref{atlas}, 
was designed primarily for precise measurements
in the highest-energy proton-proton collisions, particularly to discover
new high-mass particle states.  However, it is also a very capable detector
for the measurement of heavy ion collisions at the highest energies achieved
to date, already a factor of 14 higher than available at the Relativistic
Heavy Ion Collider.
The ATLAS inner detector is immersed in a 2 Tesla solenoid magnetic field and
provides precise reconstruction of particle trajectories typically
with three pixel layers and four double sided silicon strip detectors
out to $|\eta|=2.5$, as
well as a transition radiation tracker, out to $|\eta|=2$.
The longitudinally segmented ATLAS calorimeter provides electromagnetic
and hadronic measurements out to $|\eta|=4.9$, with particularly
high spatial precision in the $\eta$ direction.  
Finally, the ATLAS Muon Spectrometer is located outside the calorimeters
(which range out most of the hadronic backgrounds) and measures muon
tracks out to $|\eta|<2.7$.

The LHC provided lead-lead collisions to the three large experiments in
November and December 2010, and ATLAS accumulated almost ten inverse microbarns of luminosity,
as shown in the right panel of Fig.~\ref{atlas}.
The analyses shown here typically use up to eight inverse microbarns,
during which the main solenoid was activated.  A smaller dataset with
field off was used for multiplicity analyses.

\section{Global observables}

\begin{figure}[t]
\begin{center}
\includegraphics[width=24pc]{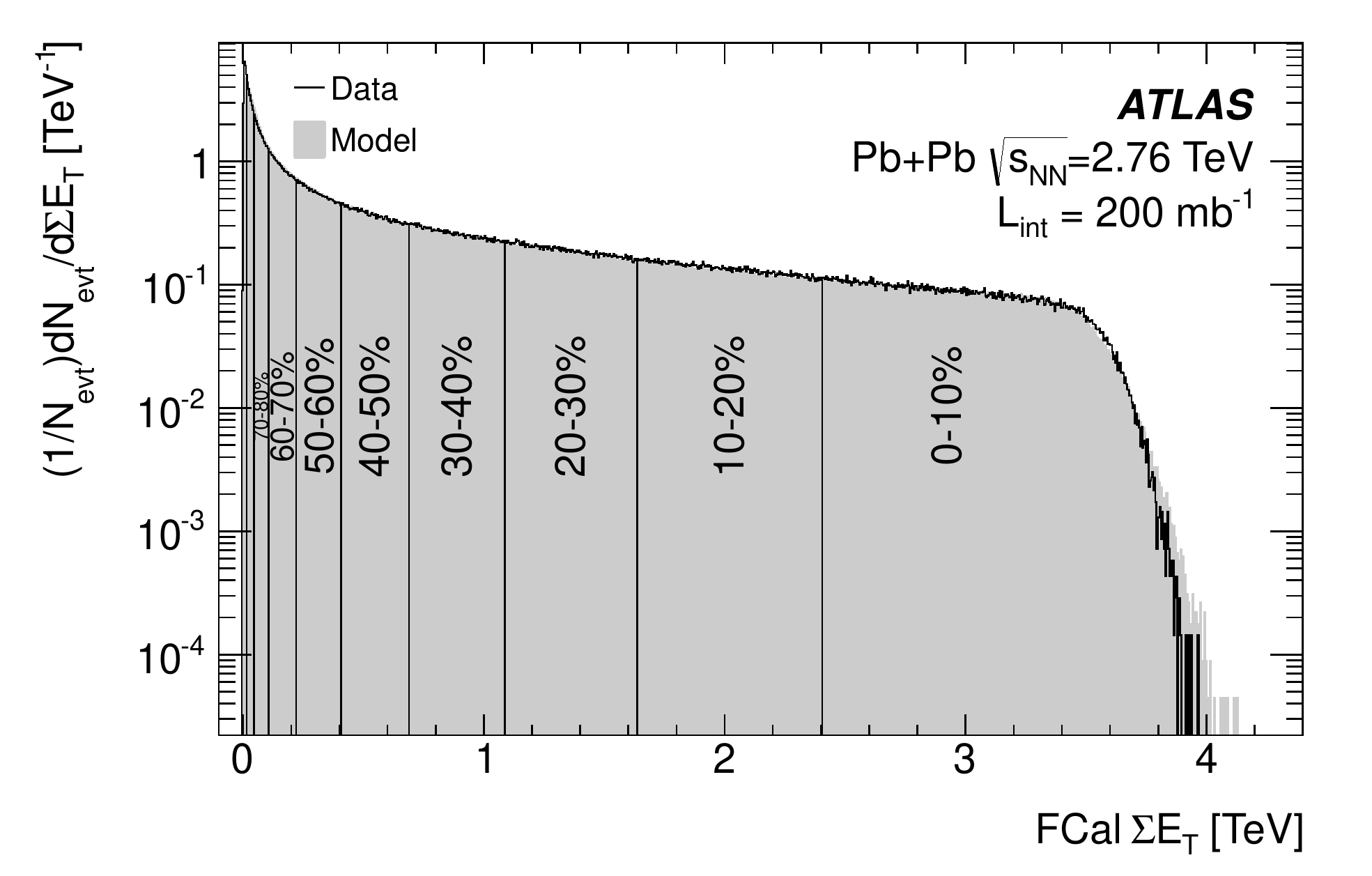}
\caption{\label{paper_centbins}
Measured distribution of FCal $\Sigma E_{\mathrm{T}}$ with percentile
bins indicated.
}
\end{center}
\end{figure}

\begin{figure}[t]
\begin{center}
\includegraphics[width=26pc]{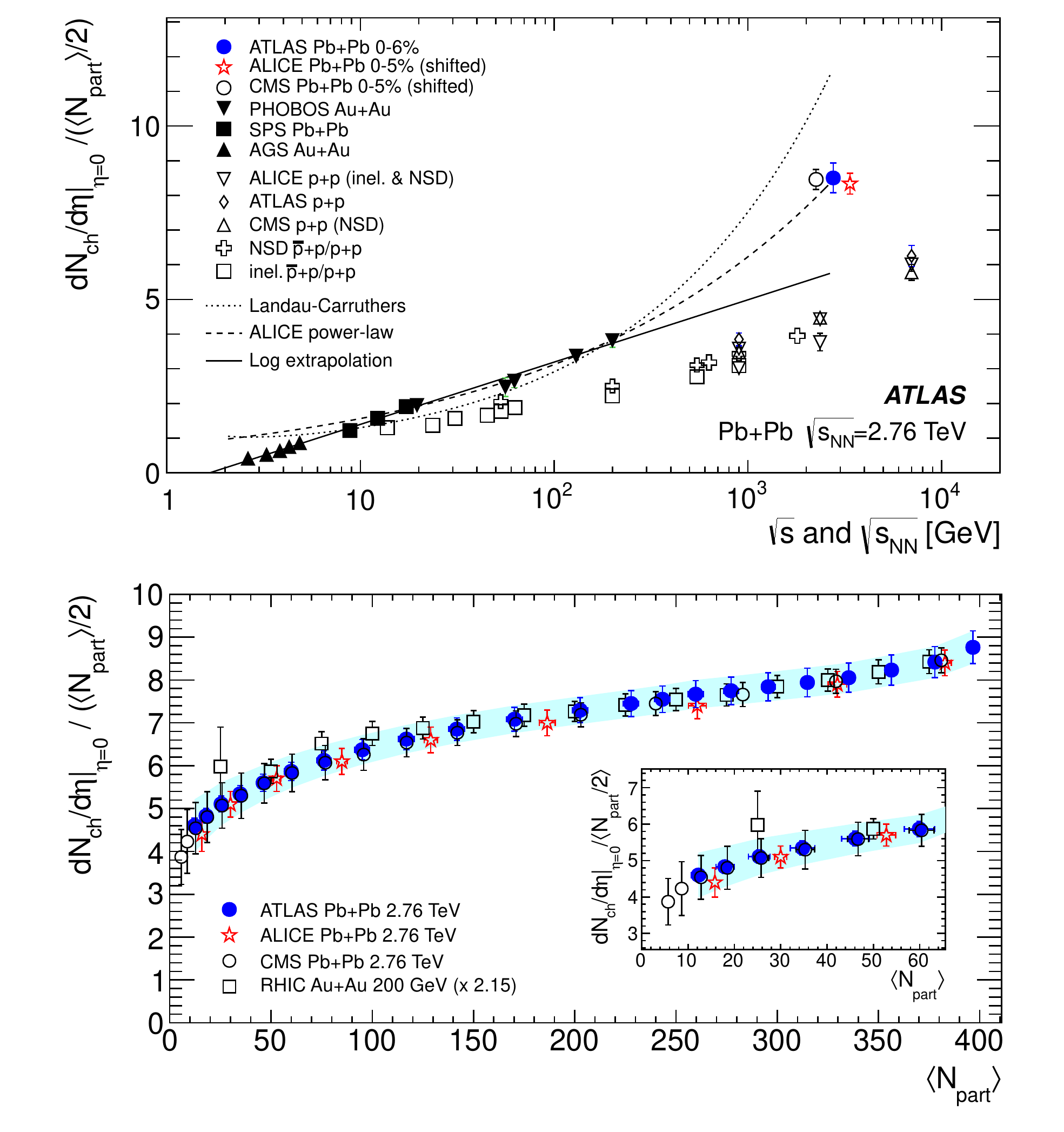}
\caption{\label{paper_dndetaNpart}
(top) Energy dependence of the charged-particle multiplicity density 
averaged over $|\eta|<0.5$ for the 0-6\% most central events compared
with heavy ion and proton-(anti)proton data~\cite{QM2011Mult}.
(bottom) The final centrality dependence of the multiplicity per participant
pair~\cite{QM2011Mult}, compared with ALICE and CMS data and an average of the RHIC data.
}
\end{center}
\end{figure}

Global observables, such as the total multiplicity and elliptic flow,
address crucial issues related to the basic
properties of the hot, dense medium.  
The charged particle multiplicity is important both from a thermodynamic
standpoint, as it is proportional to the initial state entropy, as well
as from the perspective of energy loss calculations which require an
estimate of the initial gluon density.
In ATLAS, the multiplicity has been measured~\cite{QM2011Mult} with the solenoid field turned
off to mitigate the losses from the bending of low momentum particles.
The pixel layers within $|\eta|<2$ are used to minimize the non-primary
backgrounds by only using hits consistent with a minimum-ionizing
particle emanating from
the primary vertex.  The pixel hits are assembled both into ``tracklets'' (two
hits, both consistent with the measured primary vertex) and full
tracks (three points, reconstructed with the main ATLAS tracking software).
In all, three methods are used, each with quite different systematic uncertainties,
providing accurate measurements up to the highest particle densities
and uncertainties of only a few percent.
The collision centrality is estimated using the total transverse energy
measured in the ATLAS forward calorimeter (FCal,
covering $3.2 < |\eta| < 4.9$) and bins are selected 
using an estimated sampling fraction of $f=100\pm2 \%$ in the preliminary results
shown at Quark Matter 2011, 
and a revised final value of $f=98\pm2\%$ in recently-submitted papers~\cite{QM2011Mult,QM2011flow}.
The final centrality selections~\cite{QM2011flow} are shown in Fig.~\ref{paper_centbins}.

The final multiplicity results are shown in Fig.~\ref{paper_dndetaNpart}.
The charged-particle yield, averaged over $|\eta|<0.5$, $dN_{\mathrm{ch}}/d\eta / \langle N_{part}/2\rangle$
for the 6\% most central events,
agrees well (consistent well within the stated systematic uncertainties) 
with the previously-available ALICE and CMS data~\cite{Collaboration:2010cz}.
There is also good agreement for all centrality intervals (parametrized
by $N_{part}$), even for the most peripheral events 
(as shown in detail in the inset).
After scaling the RHIC data by the observed factor of
2.15, the centrality dependence
at the LHC is also seen to be quite similar to that 
observed at a much lower energy.
A similar observation was made at RHIC, by comparing data from $\sqrt{s_{NN}}=200$ GeV to that from 19.6 GeV 
and finding a similar centrality dependence there as well.
Given the dramatic increase in the rates of hard processes at higher energies, 
this seems to suggest that the centrality dependence of the charged particle density 
is primarily geometric in its origin.

\begin{figure}[t]
\begin{center}
\includegraphics[width=30pc]{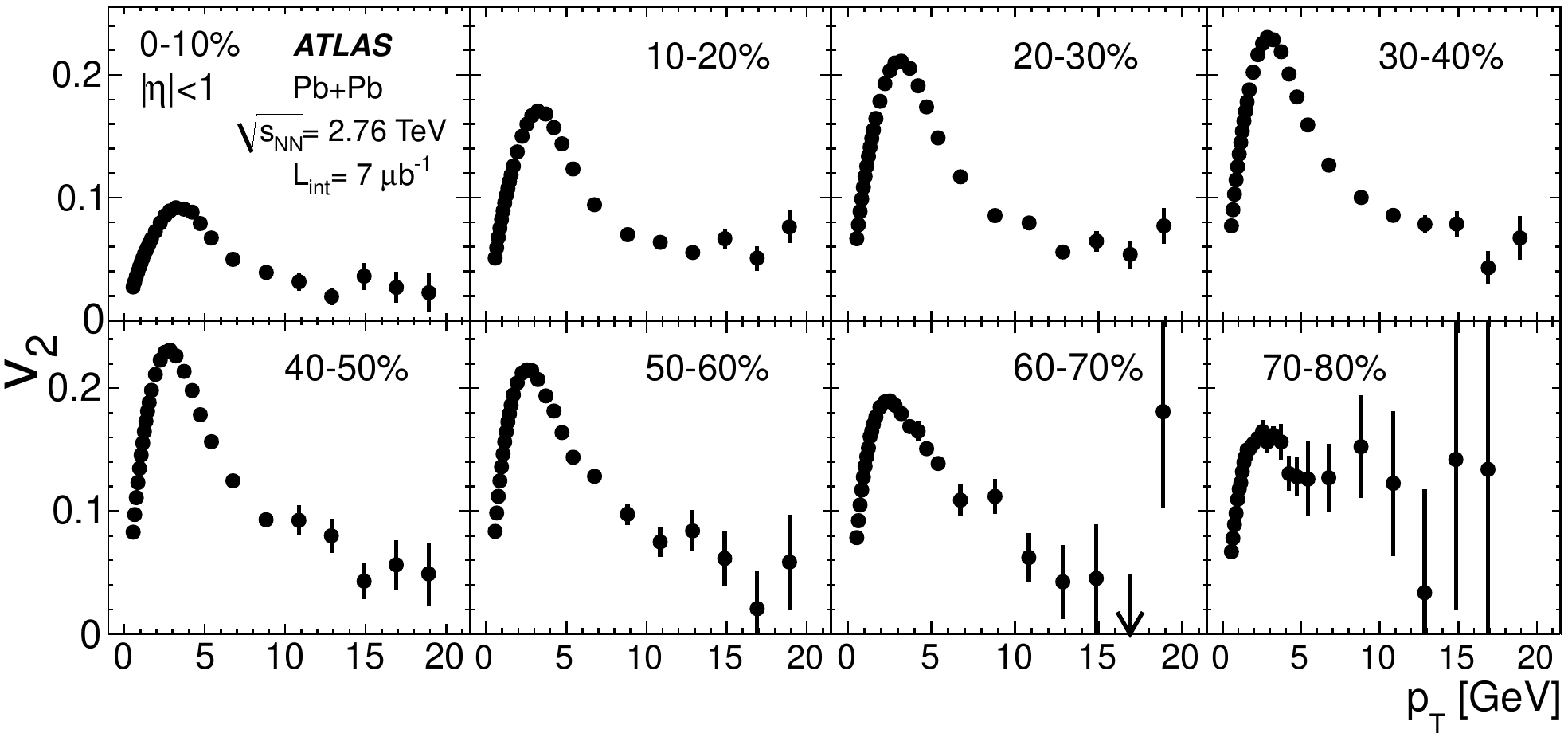}
\includegraphics[width=28pc]{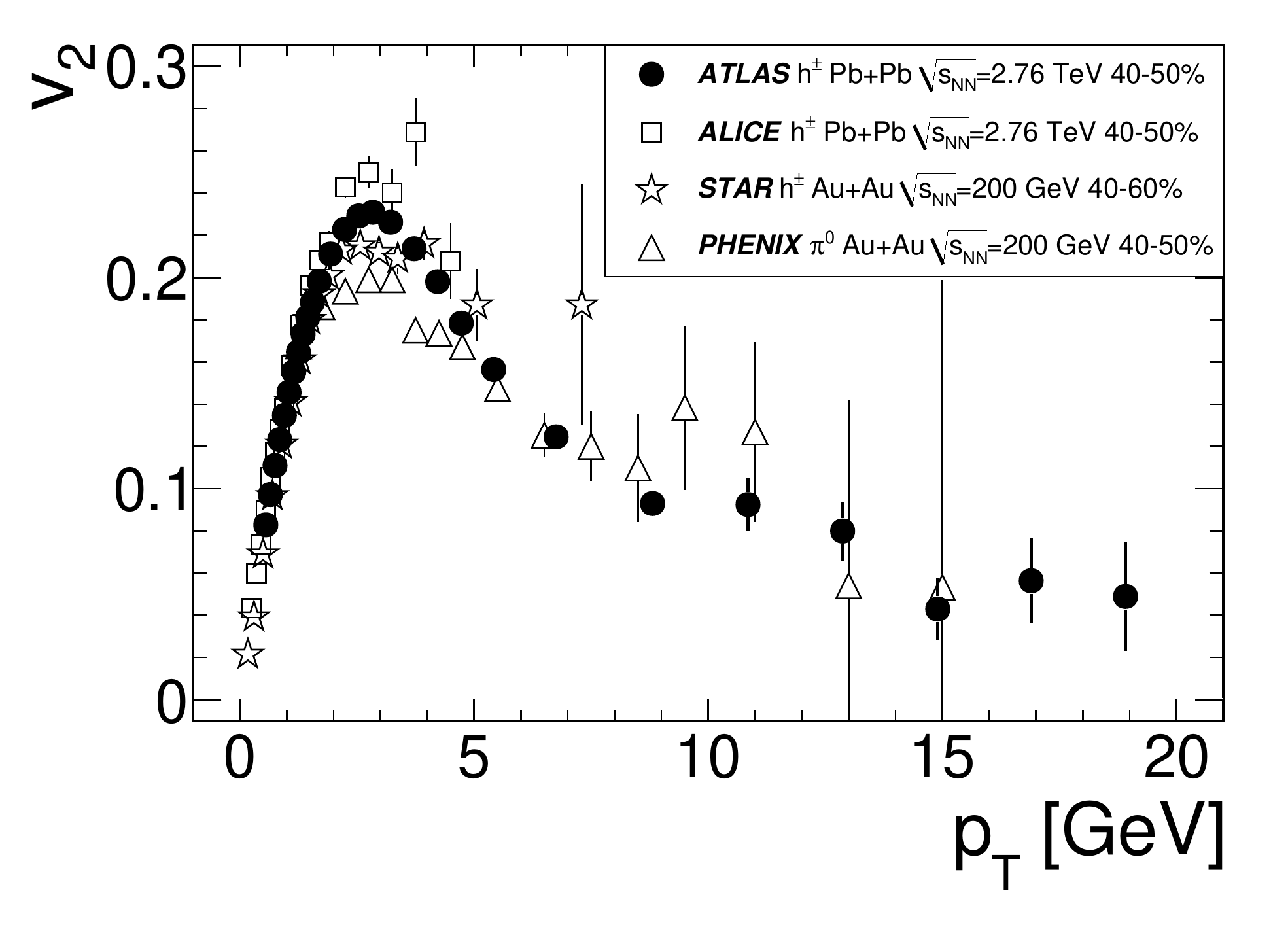}
\caption{\label{v2_vs_pt}
(top) Elliptic flow as a function of transverse momentum integrated over
$|\eta|<1$ for eight centrality intervals
(bottom) Elliptic flow as a function of transverse momentum for the 40-50\% centrality interval (\cite{QM2011flow} and references therein), compared with data from RHIC and the LHC.
}
\end{center}
\end{figure}

The collective response of the system to the initial geometric configuration
of the nucleons in the colliding nuclei is typically characterized by 
the ``elliptic flow'', or a $\cos(2[\phi-\Psi_2])$ modulation in the 
azimuthal angular distributions
relative to the measured ``event plane'' angle $\Psi_2$.
In ATLAS, the event plane is measured in the FCal. 
Elliptic flow is measured using tracks in the inner detector 
out to $|\eta|=2.5$.  The effects of ``non flow'' are minimized by
maximizing the gap between the track and the event plane measurement by using
the event plane from the opposite FCal hemisphere 
to the measured track~\cite{QM2011flow}.
The half-amplitude of the elliptic flow harmonic ($v_2$), 
as a function of transverse momentum and pseudorapidity,
has been measured out to $\pT=20$ GeV and $|\eta|<2.5$ for eight centrality bins.
There is no substantial pseudorapidity dependence out to the maximum measured
$\eta$, but there is a strong $\pT$ dependence with a pronounced rise in $v_2$ out
to $\pT=3$ GeV, a slower decrease out to $\pT=8$ GeV and then a very weak
dependence out to the highest measured $\pT$.
The top panels of Fig.~\ref{v2_vs_pt} shows the final ATLAS data on 
$v_2$ vs. $\pT$ in eight centrality bins.
The bottom panel shows the ATLAS data from the 40-50\% interval
compared with already-published ALICE~\cite{alicepaper} data 
as well as lower energy data from PHENIX~\cite{Adare:2010sp} and 
STAR~\cite{Adams:2004bi}, 
all from the 40-50\% centrality bin (except STAR data from 40-60\%).  
One finds that all of the data are quite similar, even at high $\pT$
within the large statistical errors of the PHENIX $\pi^0$ data.

\begin{figure}[t]
\begin{center}
\includegraphics[width=16pc]{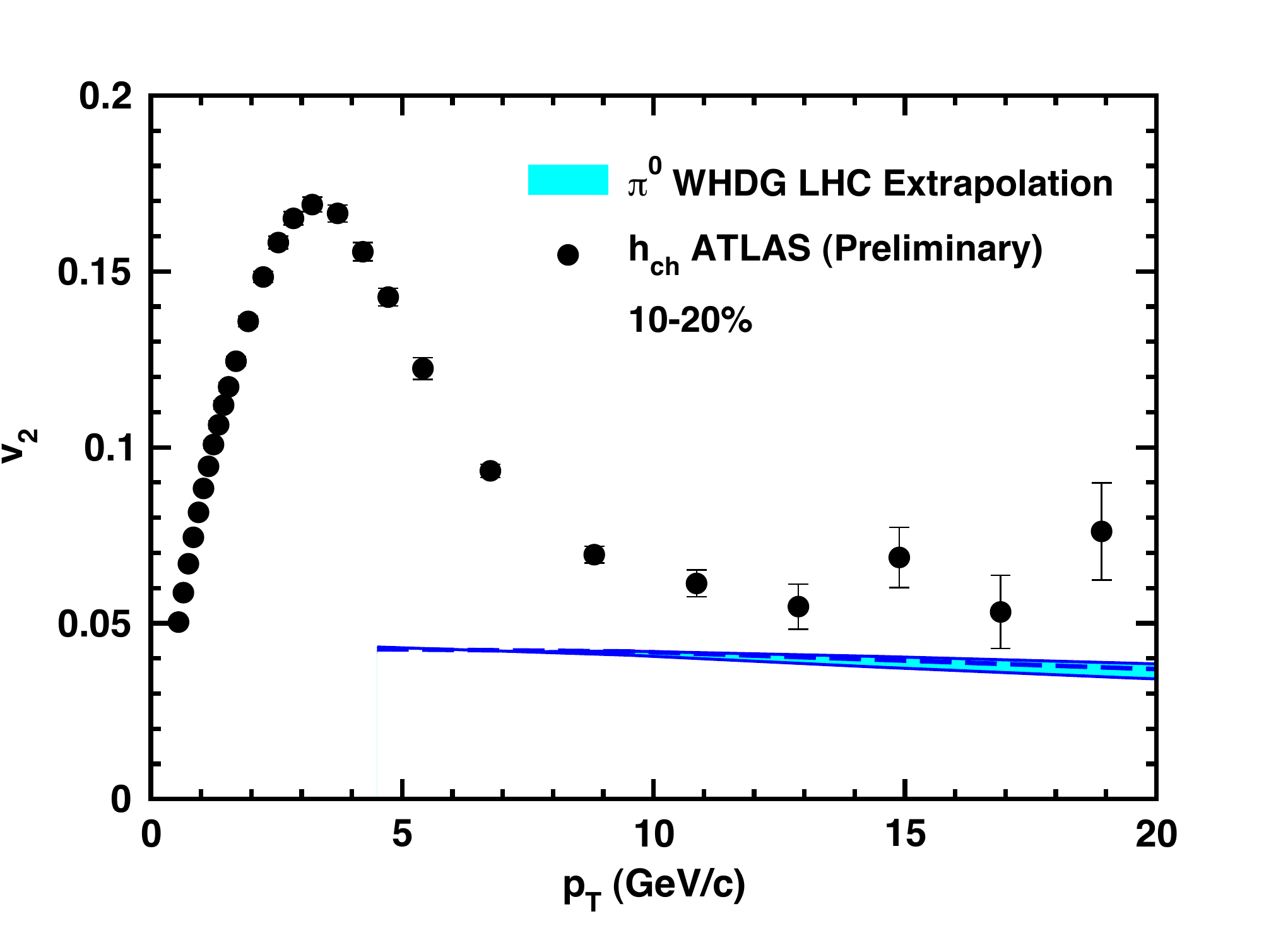}
\includegraphics[width=16pc]{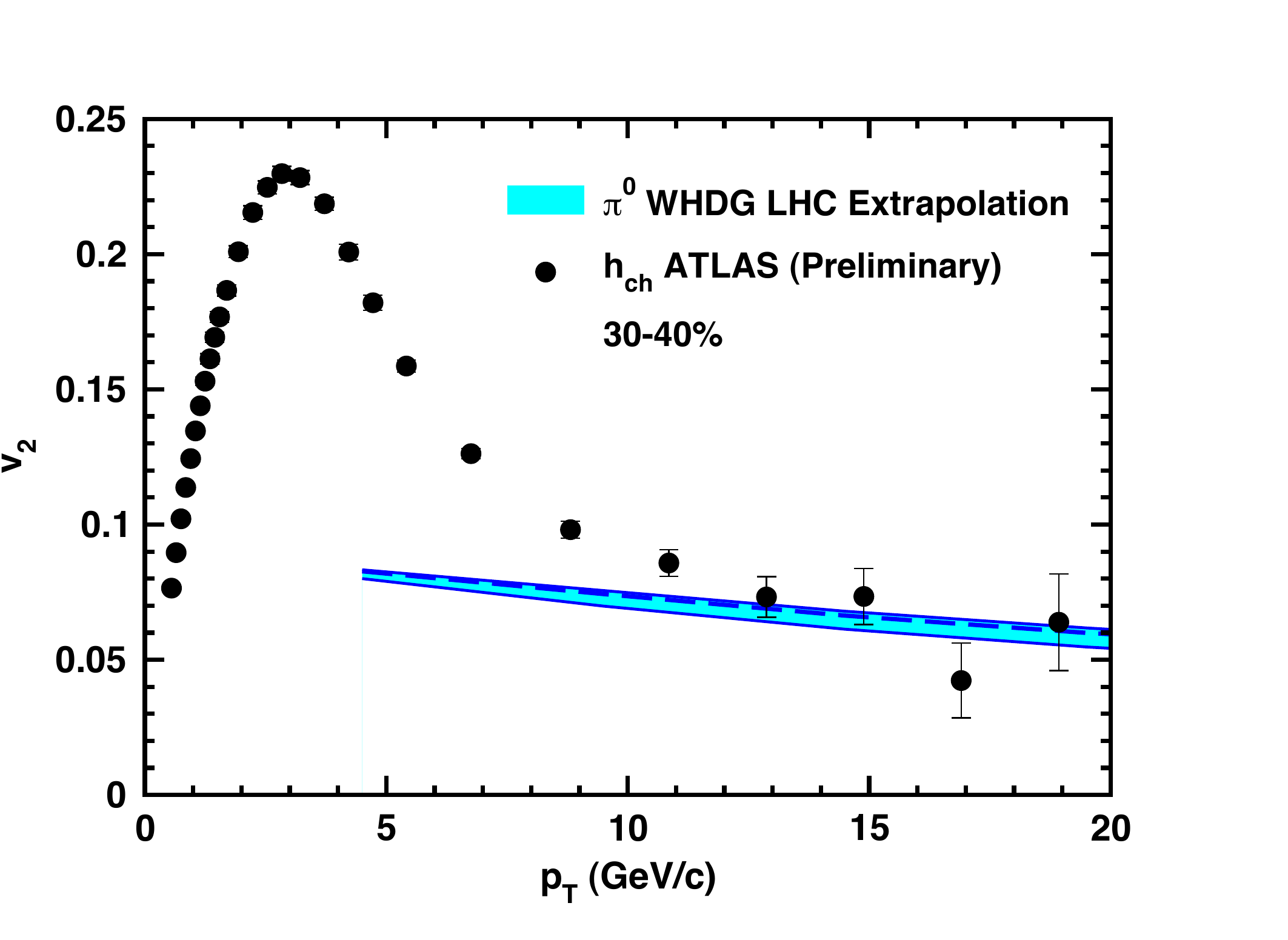}
\includegraphics[width=16pc]{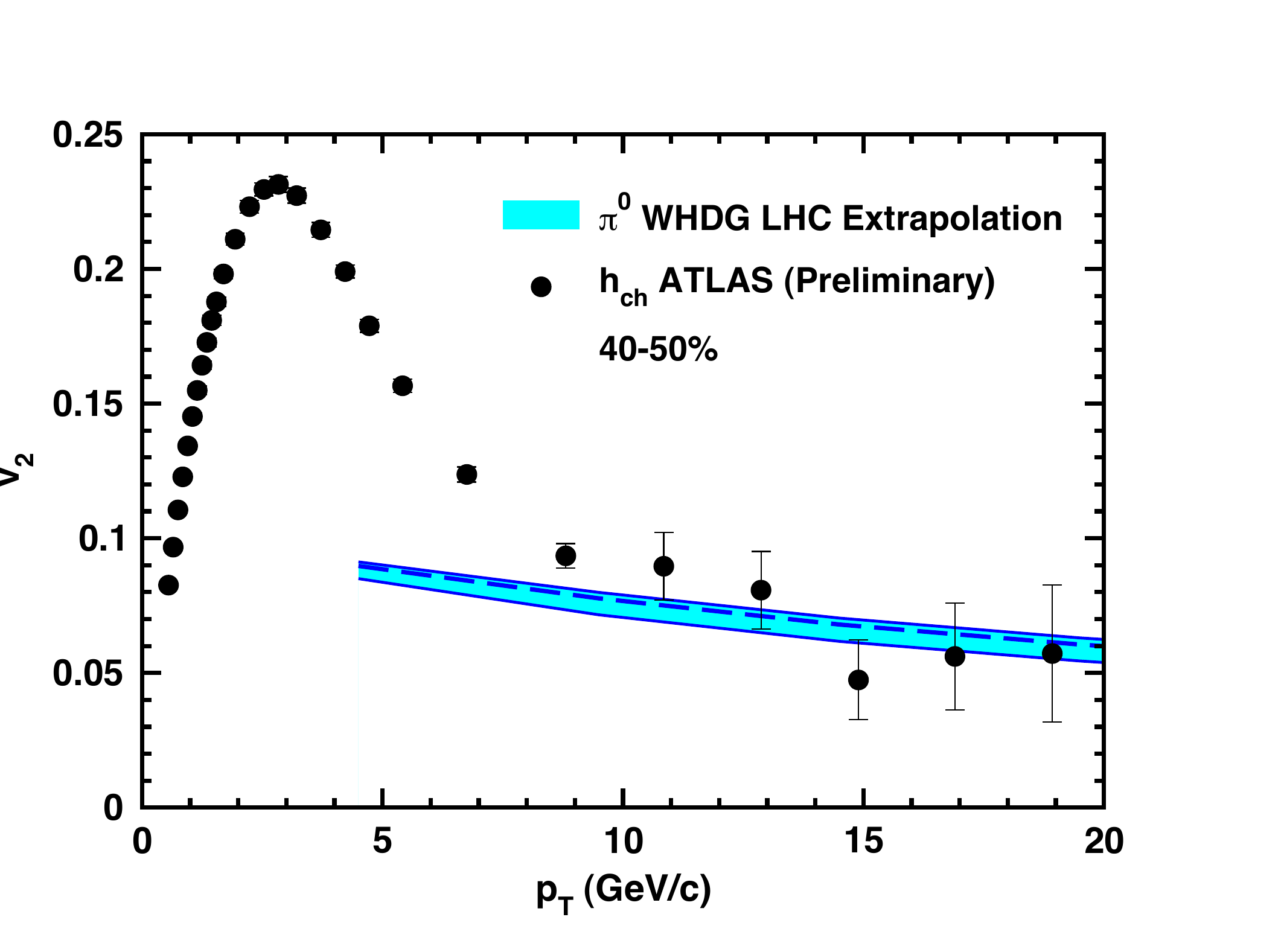}
\caption{\label{fig:horowitz}
Comparisons of ATLAS data of $v_2$ at high $p_T$ compared with differential energy loss.
}
\end{center}
\end{figure}

It is not obvious if this apparent scaling behavior, especially at high $\pT$,
is consistent with expectations from jet energy loss.
A first look at $v_2$ at high $\pT$ from energy loss calculations 
has been made by by Horowitz and Gyulassy~\cite{horowitz} 
and is shown in Fig.~\ref{fig:horowitz}.
These calculations are found to describe high $\pT$ hadron suppression data at RHIC quite well.
An extrapolation to LHC energies is performed by scaling the initial gluon density by the 
ratios of the charged-particle multiplicity near $\eta=0$.
However, it is found that the hadron suppression thus predicted at the 
LHC is a factor of two smaller than is observed in the recent LHC data 
(i.e. the measured $R_{AA}$ is a factor of two higher than predicted).  
The predicted values of $v_2$ reflect the differential between energy
loss of jets emitted ``in plane'' (i.e. which travel a shorter path length 
through the medium, and thus are less suppressed) and ``out of plane'' 
(which are conversely more suppressed).
When compared to the preliminary ATLAS data, it is found that the data agree 
surprisingly well with the predictions for $\pT>10$ GeV and centrality 
above 30\%.  The more central data (10-20\%) show a sigificant divergence 
from the predictions.  
However, this may be explained in part~\cite{horowitzp} by 
the lack of fluctuations in the initial nucleon configurations used in the 
calculations~\cite{Horowitz:2011gd}.

\begin{figure}[t]
\begin{center}
\raisebox{3mm}{\includegraphics[width=20pc]{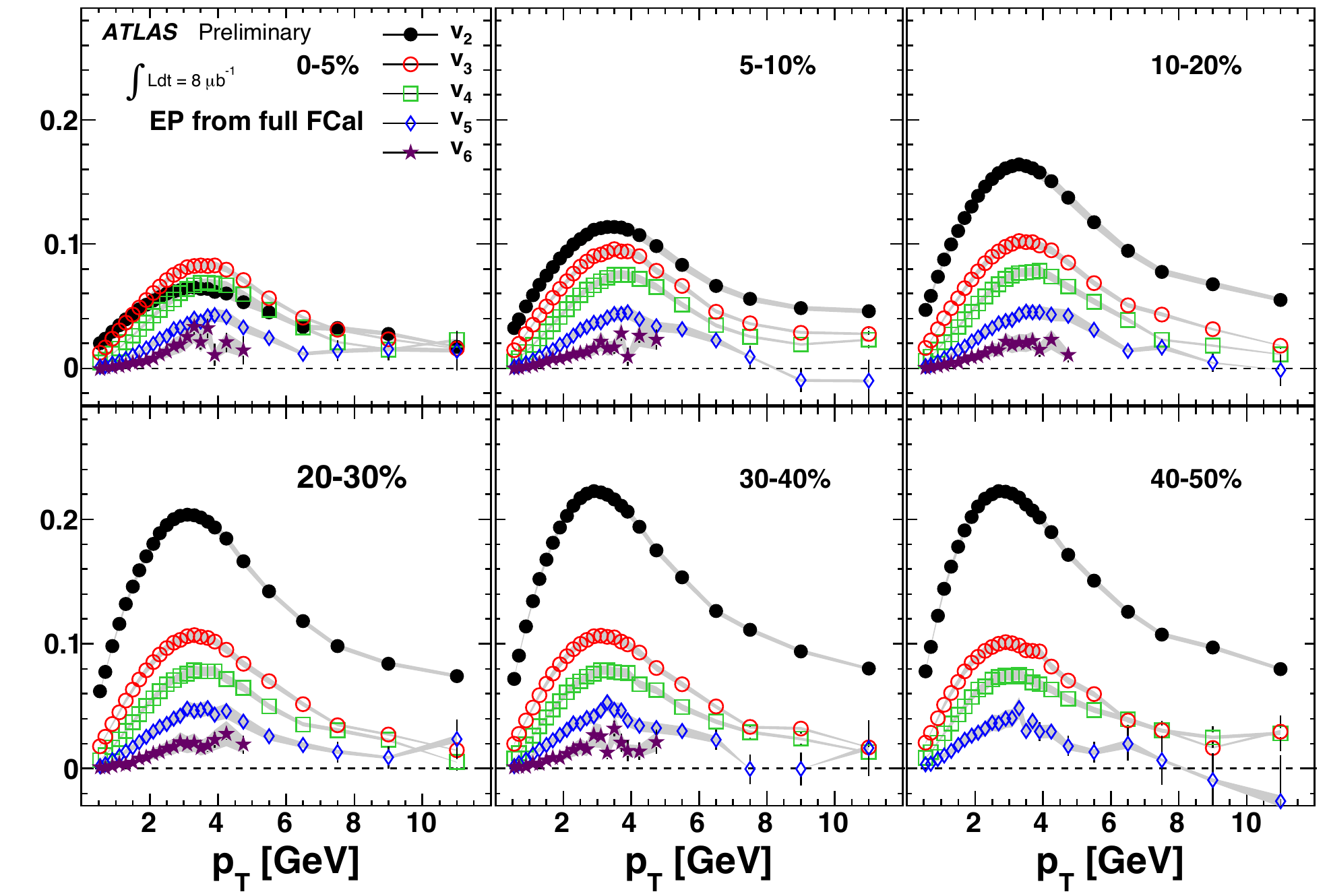}}
\includegraphics[width=20pc]{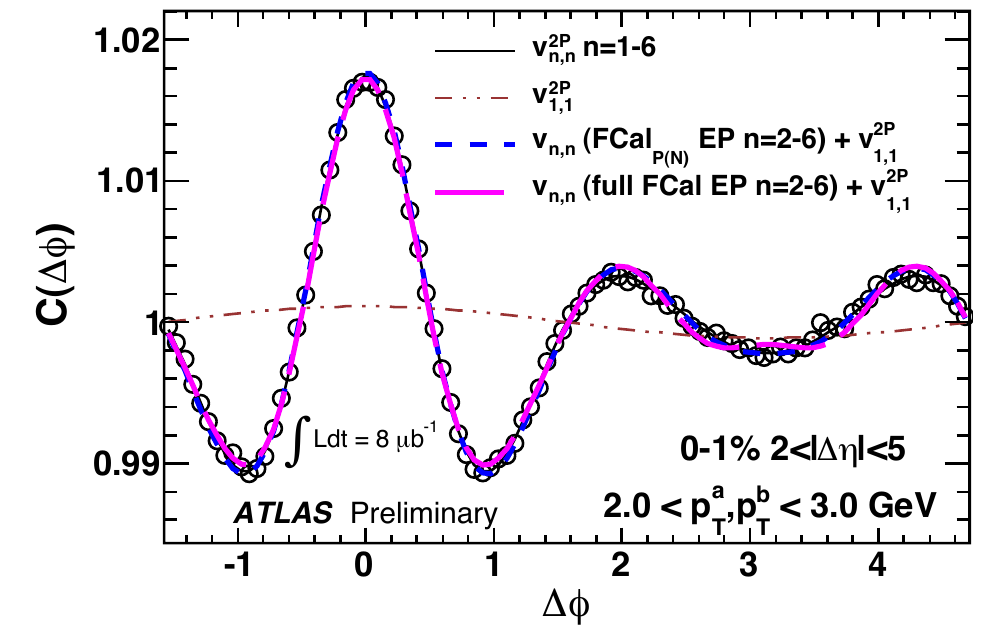}
\caption{\label{2pc}
(left) Higher order Fourier coefficients ($v_2-v_6$) as a function of transverse
momentum for six centrality bins, using the Event Plane (EP) method.  
It is observed that while $v_2$ varies
strongly with centrality, $v_3-v_6$ are nearly invariant.  
(right) Measured two-particle correlation function for $2<\pT < 3$ GeV 
and $\eta < 2.5$ in the 1\% most central collisions (open circles).
The reconstructed correlation function using the coefficients from
the event plane measurement (and $v_1$ from the two-particle measurement)
is also shown and agrees quite well with the directly measured correlation
function, using both full FCal (solid line) and FCal$_\mathrm{P(N)}$ (dashed line) methods.
}
\end{center}
\end{figure}

After the realization that initial state fluctuations can induce 
higher Fourier modes in the overlap region of the two nuclei~\cite{Alver:2010gr}, elliptic
flow has now been joined by the study of higher-order harmonic
flow coefficients~\cite{QM2011fourier}.  These are measured each in their own event plane.
The resolution has been evaluated by a variety of subdetector combinations
and it is found that ATLAS can resolve modes up to $n=6$ in the most
central 40\% of events.
These are shown as a function of $\pT$ for 
six centrality bins in the left panel of Fig.~\ref{2pc}.
These figures show the rapid change of $v_2$
with centrality, expected from previous measurements and attributed to the
changing overlap geometry as the impact parameter is increased. 
They also show only a modest change in the higher order
coefficients as the centrality changes, suggesting that they are mainly sensitive to 
fluctuations of nucleon positions relative to the overall elliptical shape given by
$v_2$.
Hydrodynamic calculations suggest a strong dependence of $v_n$ on $n$, particularly when
viscous effects are taken into consideration.
Thus, it is expected that the information from these higher coefficients
can distinguish between competing physics scenarios attempting to 
explain the initial state 
and will provide new information on the viscosity to entropy ratio.


A longstanding puzzle at RHIC is the observation of unusual structures 
(e.g. the Mach Cone~\cite{CasalderreySolana:2004qm})
in the two particle correlation function, particularly on the ``away side''
relative to a high $\pT$ particle.  
Both STAR~\cite{Adams:2005ph} and PHENIX~\cite{Adare:2007vu} observed a strongly-modified away-side 
peak which even developed a ``dip'' at $\Delta\phi = \pi$ in the most central events. 
The ``near side'' is also found to show an ``ridge''-like
enhancement at $\Delta\phi \sim 0$ at large values of $\Delta \eta$~\cite{:2009qa}.
However, both of these phenomena have been argued to arise most likely from 
the presence of higher 
flow coefficients beyond elliptic flow~\cite{Alver:2010gr}.
In ATLAS, the correlation function has been measured with large separation in $\eta$ to
suppress the contribution from jets and a discrete Fourier transform (DFT)
is used to extract coefficients out to $n=6$~\cite{QM2011flow}.  These are found to agree
quite well with the $v_n$ extracted using event plane approaches.  More
interestingly, when the event plane coefficients are used to construct
a comparable two particle correlation function $dN/d\phi \propto \sum_n v^2_n \cos(n\phi)$, one finds a striking agreement with the correlations 
measured directly, as shown in Fig.~\ref{2pc}.  
This suggests that what were formerly thought to be indications of
jet-medium interactions may well result simply from the presence of 
higher-order flow harmonics, arising through
a combination of nucleon position fluctuations and viscous effects.

\section{Hard Probes}

\begin{figure}[t]
\begin{center}
\includegraphics[width=18pc]{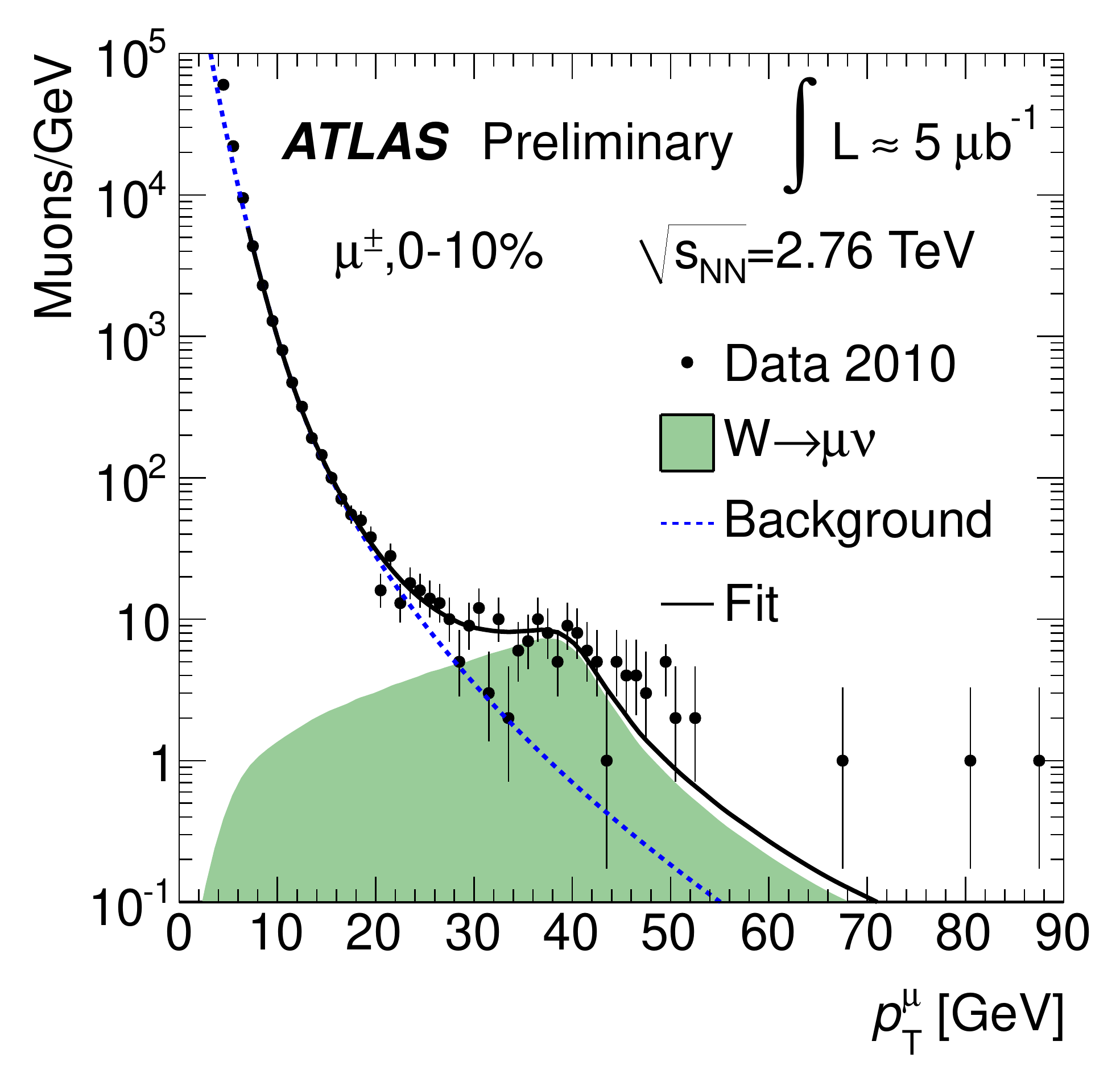}
\includegraphics[width=18pc]{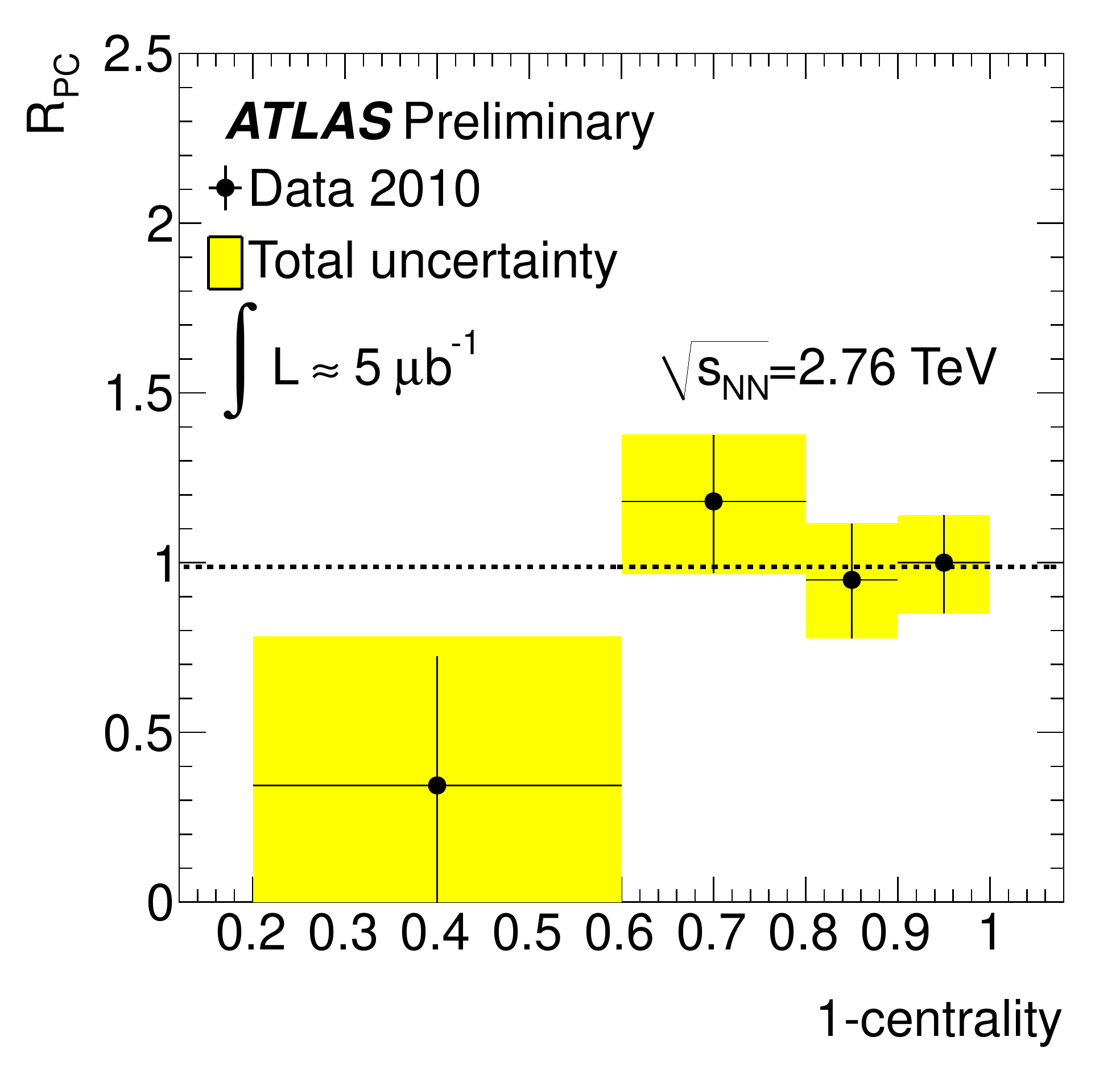}
\caption{\label{W}
(left) Single muon spectrum measured for the 0-10\% most central events,
with the templates for $W$ bosons and heavy flavor (indicated as ``Background'') 
also shown to illustrate the yield extraction procedure.
(right) $R_{PC}$ for $W$ bosons as a function of centrality, showing 
consistency with binary collision scaling.  The dotted line is a fit to
a constant.
}
\end{center}
\end{figure}

High $p_T$ signals produced by perturbative processes in the
initial collisions of the two nuclei (also known as ``hard probes'') 
provide a means to probe the conditions
of the collision at very early times.  This is typically done by means of ``suppression''
observables which indicate deviations from the expected scaling of yields
by the number of binary collisions, estimated using Glauber modeling
(e.g. the ATLAS $J/\psi$ measurement~\cite{:2010px}).
To confirm the assumed scaling, ATLAS has used the spectrum of single muons
at large $\pT$ to extract the yield of $W^{\pm}$ bosons, by a
simultaneous fit of a template trained on simulated $W^\pm$ as well as 
a parameterization of muons from heavy flavor quark decays, 
as shown in the left panel of Fig.~\ref{W}.
A sample of approximately 400 $W$ bosons has been extracted. 
The binary scaling of the measured yields is studied using the variable 
$R_{\mathrm{PC}}$, defined as the ratio of yields measured in different 
centrality classes to the yield measured in the 10\% most central events,
with all yields scaled by the corresponding number of binary nucleon-nucleon collisions.
The right panel of Fig.~\ref{W} shows $R_{\mathrm{PC}}$ as a function of 
centrality~\cite{QM2011W}.
Using a fit to a constant value, giving $\langle R_{\mathrm{PC}} \rangle$ = 0.99$\pm$0.10 
with a $\chi^2=3.02$ for 3 degrees of freedom, 
a significant consistency with binary scaling is observed.


\begin{figure}[t]
\begin{center}
\includegraphics[width=20pc]{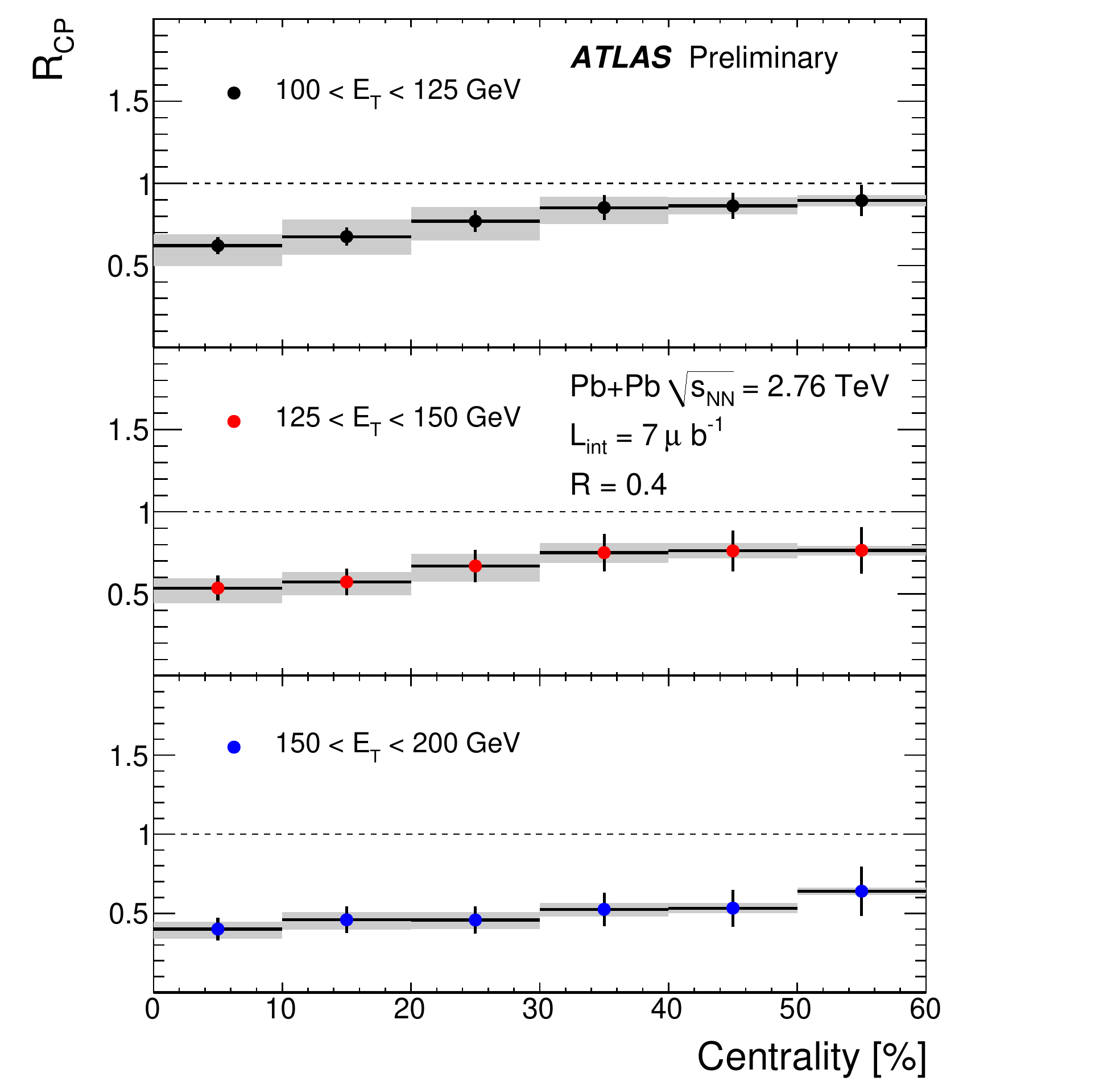}
\includegraphics[width=20pc]{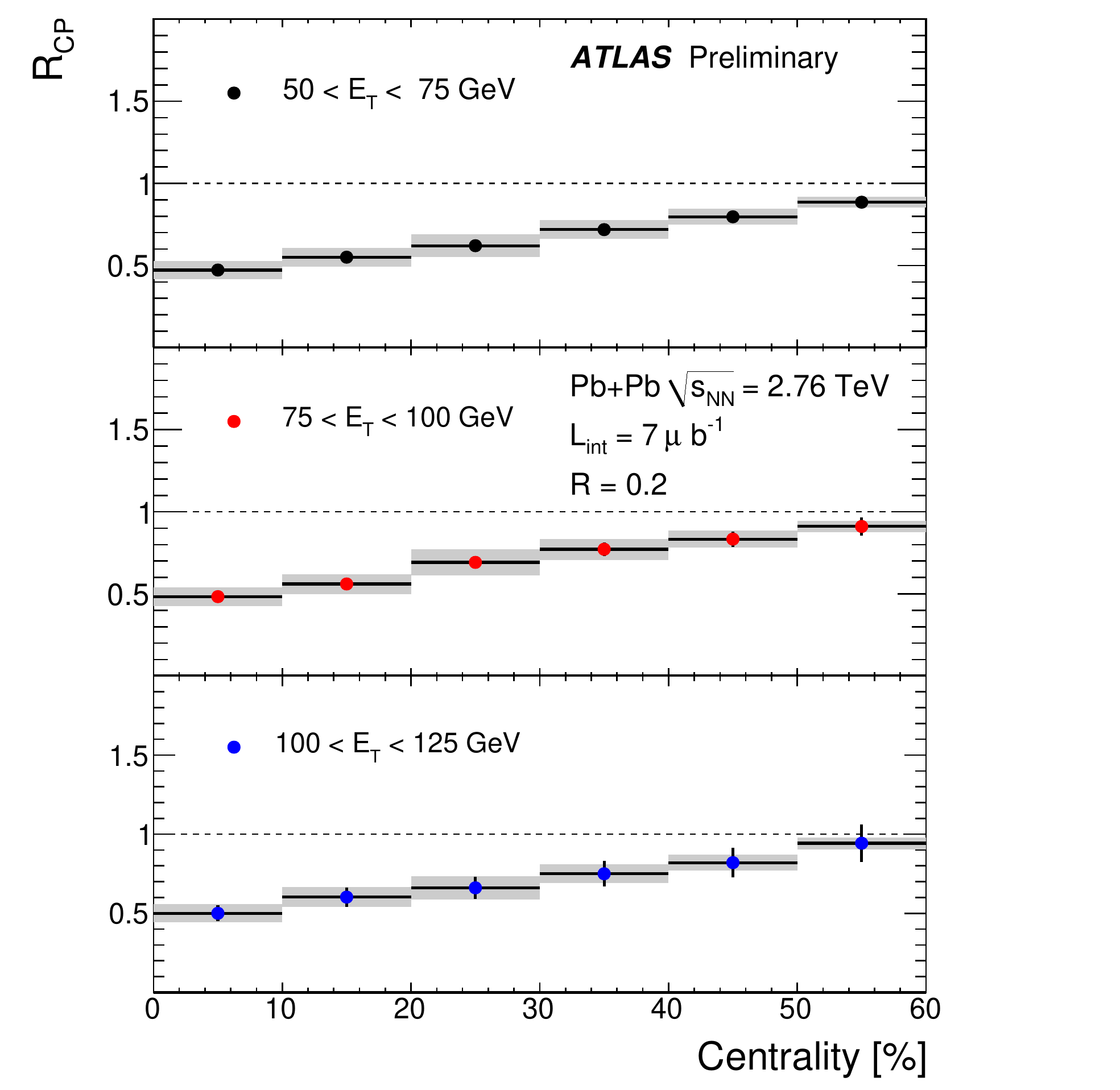}
\caption{\label{jet_rcp}
(left) $R_{CP}$ for R=0.4 jets, as a function of centrality,
with 60-80\% used as the peripheral sample.  In this convention, the most
central events are on the left and the most peripheral are on the right.
(left) Similar as the other figure, but for $R=0.2$.
}
\end{center}
\end{figure}

\begin{figure}[t]
\begin{center}
\includegraphics[width=20pc]{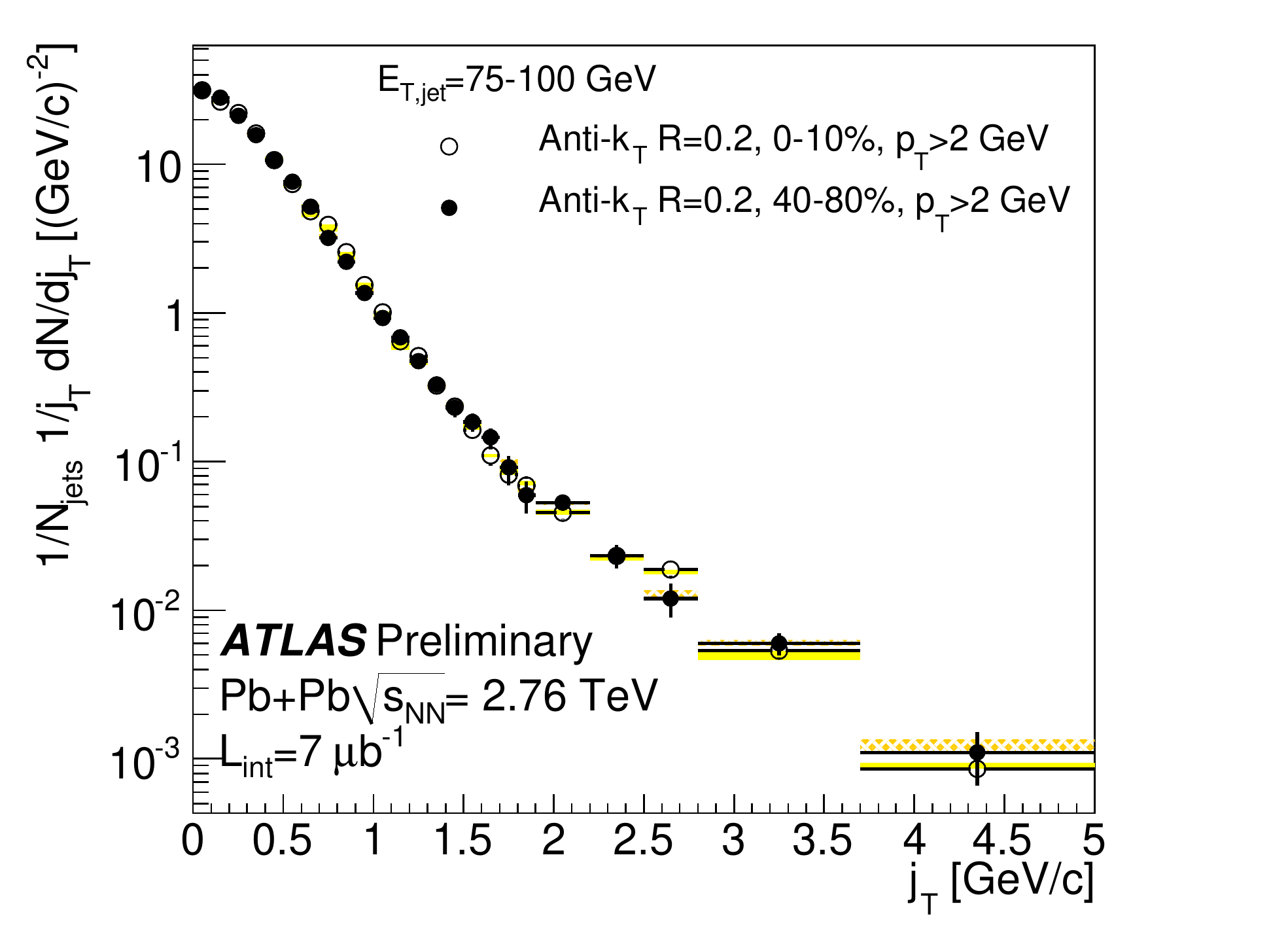}
\includegraphics[width=20pc]{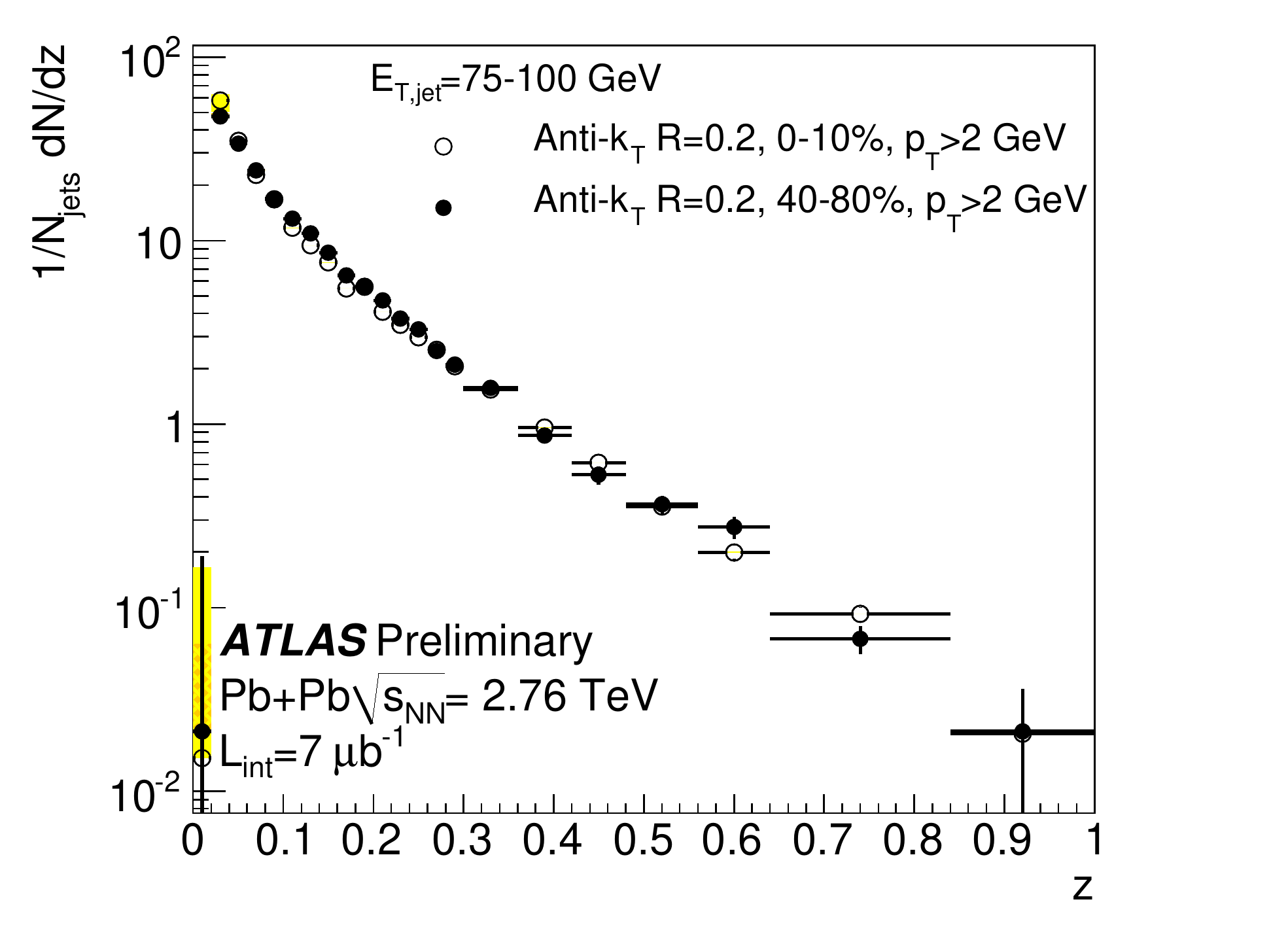}
\caption{\label{frag_func}
(left) Transverse fragmentation function for R=0.2 anti-$k_t$ jets, comparing central 0-10\% events with peripheral 40-80\%.
(right) the similar comparison for the longitudinal fragmentation function.
}
\end{center}
\end{figure}

ATLAS has already published striking results on dijet asymmetries~\cite{ATLASjetq}.
New results are reported here on the
measurement of inclusive jet yields as a function
of jet $\eT$, in order to test models of QCD energy loss more directly~\cite{QM2011jets}.
Jets are reconstructed  using the anti-$k_t$ algorithm
with the jet size set to both R=0.2 and R=0.4, based on ``towers'' composed of
calorimeter cells integrated over regions of size $\Delta\eta \times \Delta\phi = 0.1\times 0.1$.
The ambient background is removed at the cell level by excluding regions 
near jets and calculating the mean energy, as well as any azimuthal modulation,
in strips of width $\Delta\eta=0.1$.
An iterative procedure is applied to remove any residual effect of the jets
on the background subtraction.
The final jets are corrected for the energy scale
and resolution based on PYTHIA jets embedded into HIJING,
with conservative systematic uncertainties assigned to account for the
small differences between the fluctuations seen in data and simulation.
Jets are restricted to $|\eta|<2.8$, to stay within
the main barrel and endcap regions of the calorimeter.

The left panel of Figure~\ref{jet_rcp} shows $R_{\mathrm{CP}}$ for R=0.4 jets,
(defined similarly to $R_{\mathrm{PC}}$,
but relative to the 60-80\% centrality interval) as a function of centrality for jets
in three fixed $\eT$ bins ($\eT = 50-75$ GeV, $75-100$ GeV and $100-125$ GeV).
A smooth evolution of $R_{\mathrm{CP}}$ is observed going from the 50-60\% central events
on the right to the 0-10\% central events on the left, where a suppression of
roughly a factor of two is observed for all $\eT$ bins and for both R=0.4 (left panel)
as well as R=0.2 (shown the right panel).  
The invariance with R is suprising given the general 
tendency of radiative calculations to predict a modified fragmentation function
for quenched jets, with substantial transverse broadening.
However, direct measurements of 
the transverse and longitudinal fragmentation functions
for tracks with $\pT>2$ GeV, 
shown in the left and right panels respectively of Fig.~\ref{frag_func}, 
confirm that no substantial modification of the
fragmentation function can be observed when comparing peripheral and central events.
In other words, the increased suppression of the jet rates
is not accompanied by any evident modification of the jets themselves.

\begin{figure}[t]
\begin{center}
\includegraphics[width=24pc]{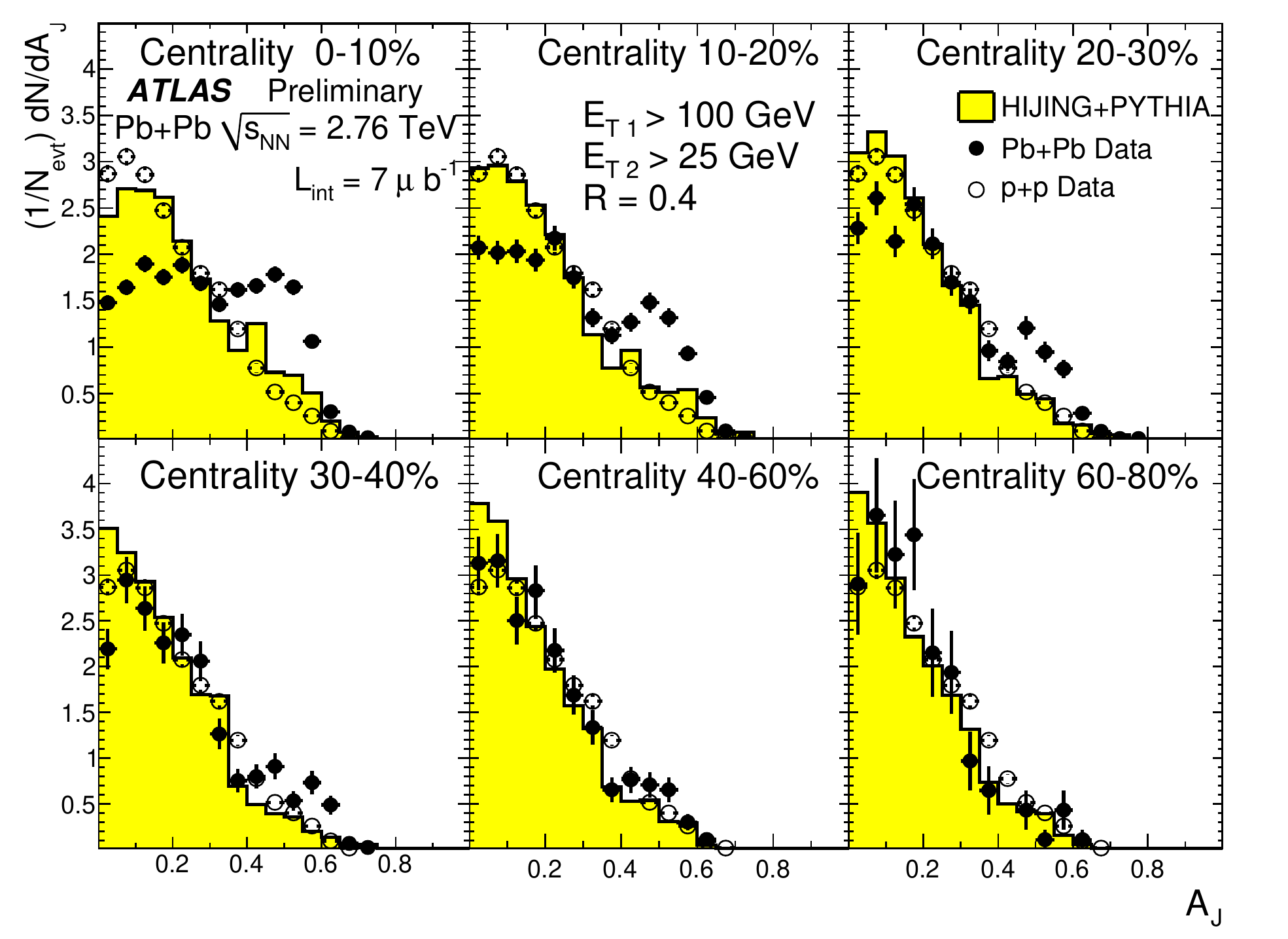}
\includegraphics[width=24pc]{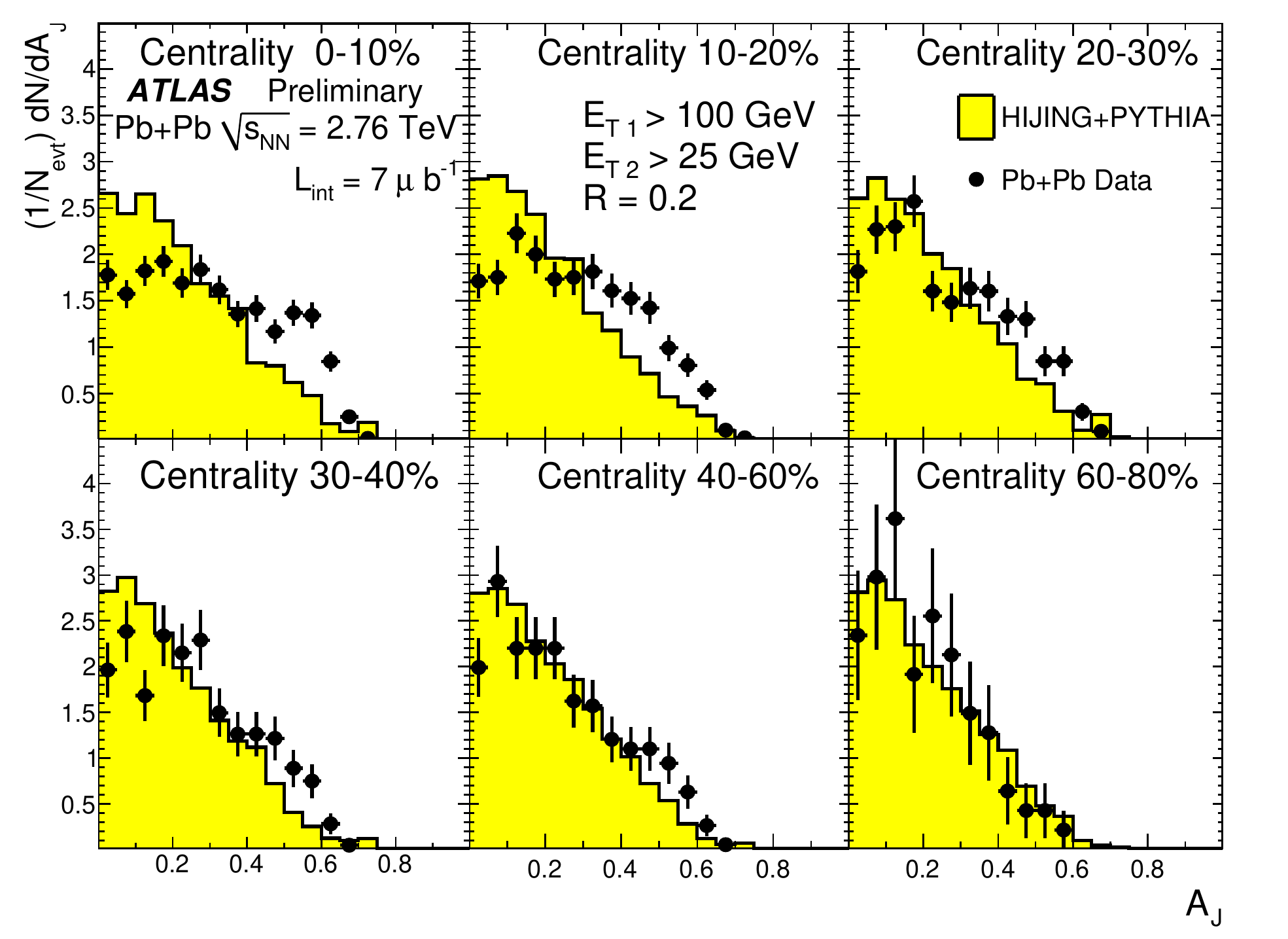}
\includegraphics[width=24pc]{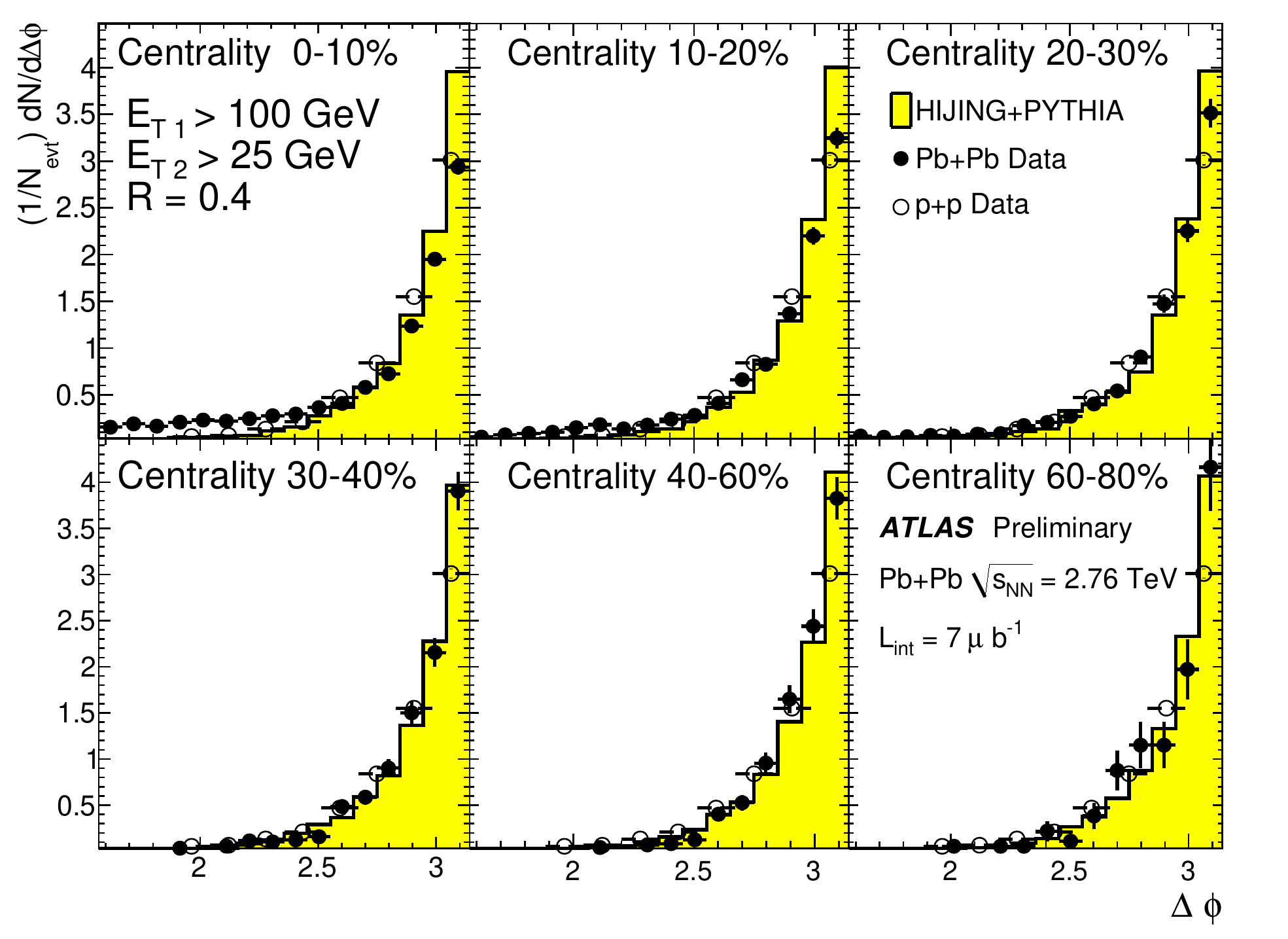}
\caption{\label{jet_asym}
(top) Asymmetry distribution as a function of centrality for $R=0.4$ jets
(middle) Asymmetry distribution as a function of centrality for $R=0.2$ jets
(bottom) $\Delta \phi$ distribution for $R=0.4$ jets.
}
\end{center}
\end{figure}

The original dijet asymmetry results provided the first direct evidence that jets
lost energy through the hot, dense medium.  The preliminary updated asymmetry results
provide several major advances over the earlier ones.  The full integrated luminosity 
($L_{\mathrm{int}}=7$ $\mu$b$^{-1}$ is now used, which increases the jet statistics by more than a factor of five.
For the larger R=0.4 jets, the background subtraction now accounts for the effect of elliptic flow, using the measured event plane from the
forward calorimeter and the $\cos(2\phi)$ modulation measured in the different calorimeter layers.
Finally, a residual effect of the background subtraction on the reconstructed jet energy is removed using
an iterative procedure that performs a second background subtraction excluding all reconstructed jets 
with $\eT>50$ GeV.  This has an approximately 10\% effect on jet energies.

The new results all use a leading jet with $\eT>100$ GeV and a subleading jet with $\eT>25$ GeV, 
with both jets within $|\eta|<2.8$ and $\Delta \phi>\pi/2$.  The cut in pseudorapidity is to exclude the ATLAS
forward calorimeter, in which the centrality and reaction plane measurements are made, and the cut in
azimuth is to require the jets emerge in opposite hemispheres.
The top panel of Fig.~\ref{jet_asym} shows the asymmetry distribution for reconstructed $R=0.4$ jets,
with the elliptic flow subtraction performed, in six centrality intervals.  
It is observed that the jets measured in the most peripheral bins resemble both
fully simulated events with PYTHIA dijets embedded into HIJING background events, as well as proton-proton
data at $\sqrt{s}= 7$ TeV.  As the centrality is increased towards events with greater nuclear overlap, the
distribution measured in lead-lead data becomes broader, and becomes essentially flat for the 0-10\%
most central events.
It should be noticed that the slight ``peak'' structure seen in the original results~\cite{ATLASjetq} is no longer visible
with higher statistics and better control over the background and jet energy.
The middle panel of Fig.~\ref{jet_asym} shows the same results, but for R=0.2 jets.  The identical trend is
seen for these much smaller jets, which have a jet area a factor of four smaller relative
to the R=0.4 jets and no flow subtraction.  The fact that the asymmetry distribution similarly widens 
for the narrow jets, which are much less sensitive to residual background fluctuations, 
provides a counterargument to the one made in Ref.~\cite{Cacciari:2011tm} that 
the asymmetry distribution
could be explained by a poor understanding of the underlying event fluctuations.
The bottom panel of the figure shows the distribution of $\Delta \phi$ between leading and subleading R=0.4 jets.  
In all centrality intervals, the emission remains primarily back-to-back.  However, there is a low-lying but significant
tail at $\Delta\phi<3\pi/4$, which can presumaby 
be explained in part by the presence of fake jets.

\begin{figure}[t]
\begin{center}
\includegraphics[width=30pc]{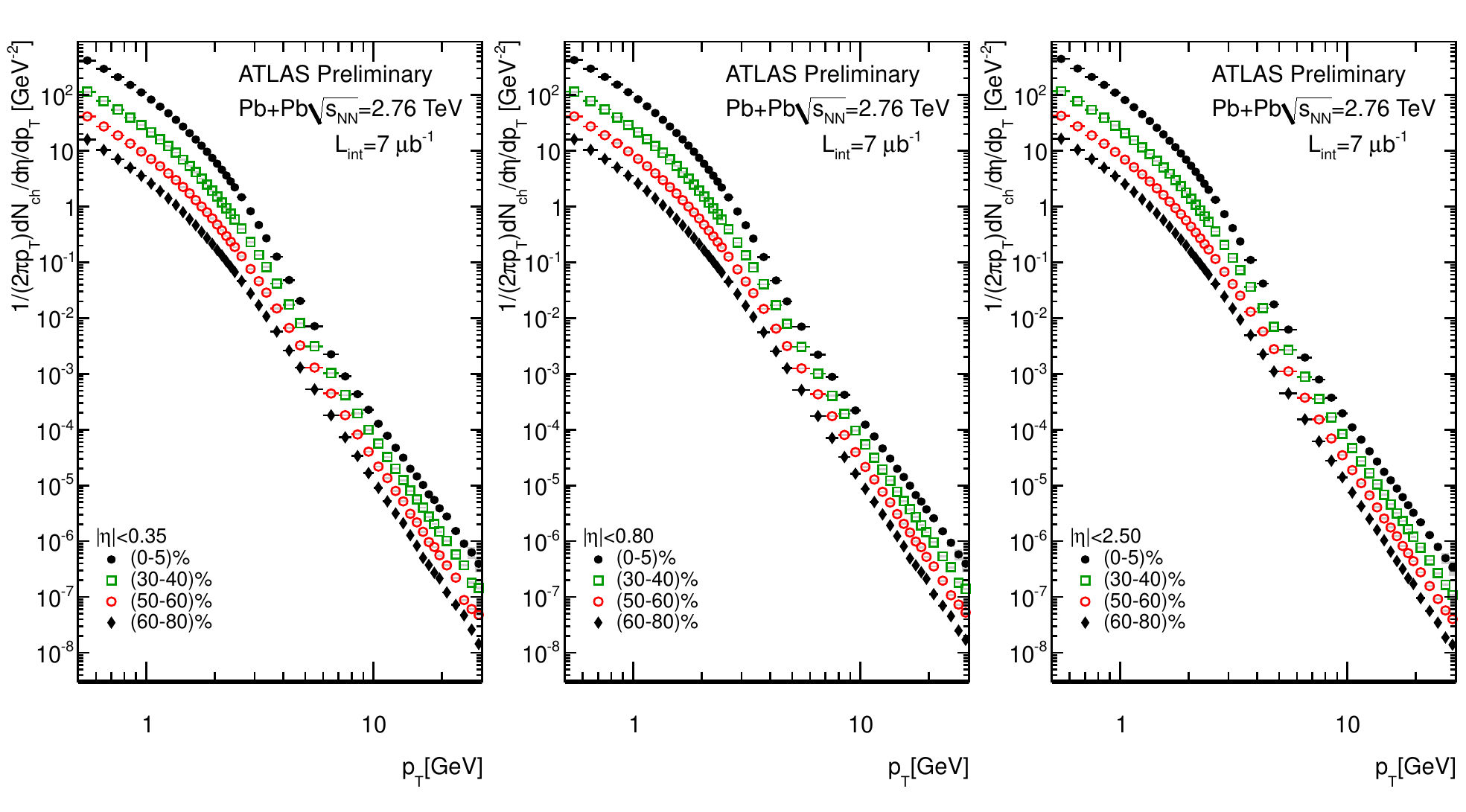}
\includegraphics[width=30pc]{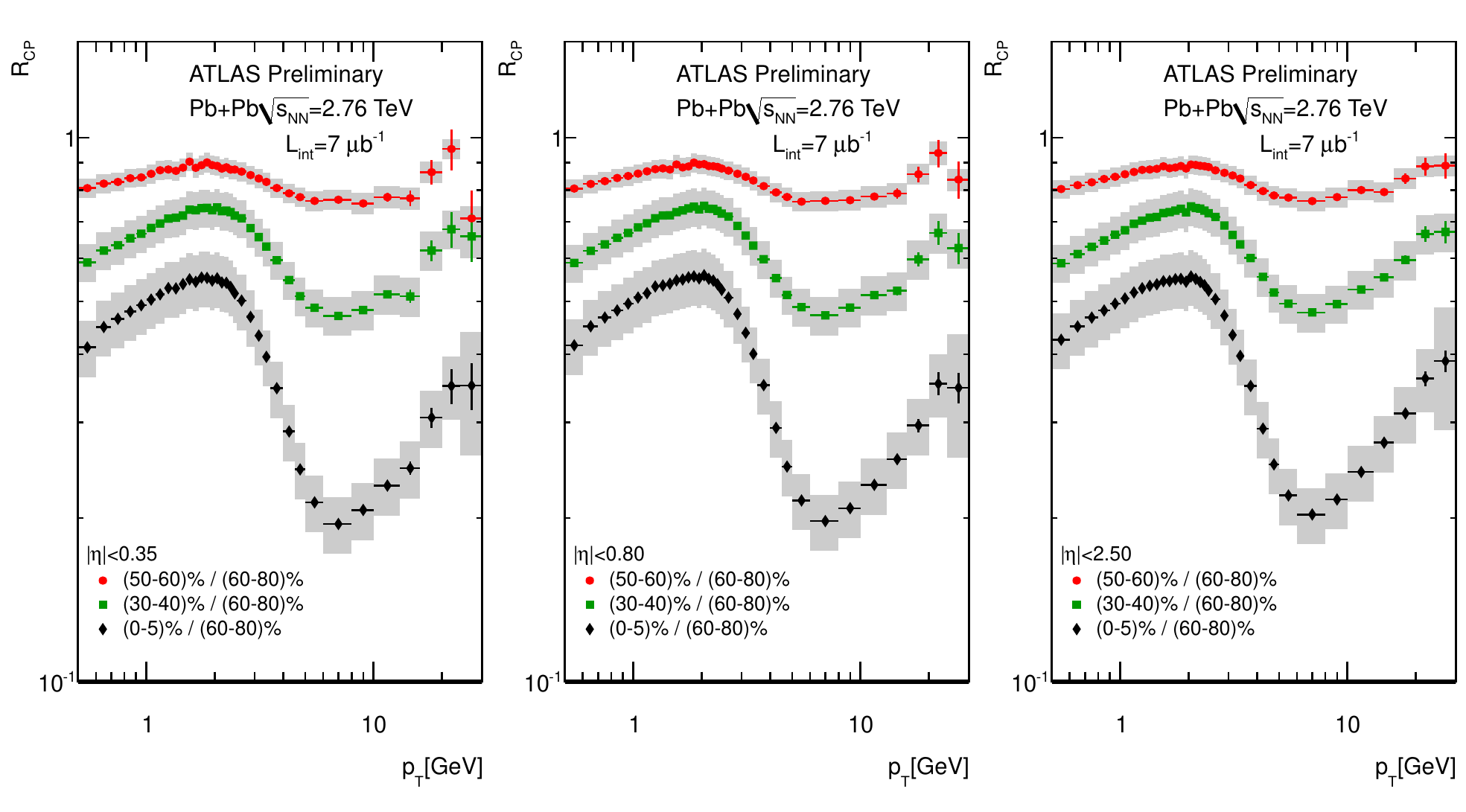}
\caption{\label{charged_rcp}
(top) Invariant yields of charged particles for four centrality intervals and 
three inclusive pseudorapidity intervals.
(bottom) Charged particle $R_{\mathrm{CP}}$ measured relative to the 60-80\% bin, for all
tracks measured within $|\eta|<0.35$, $|\eta|<0.8$ and $|\eta|<2.5$.  
}
\end{center}
\end{figure}

\begin{figure}[t]
\begin{center}
\includegraphics[width=20pc]{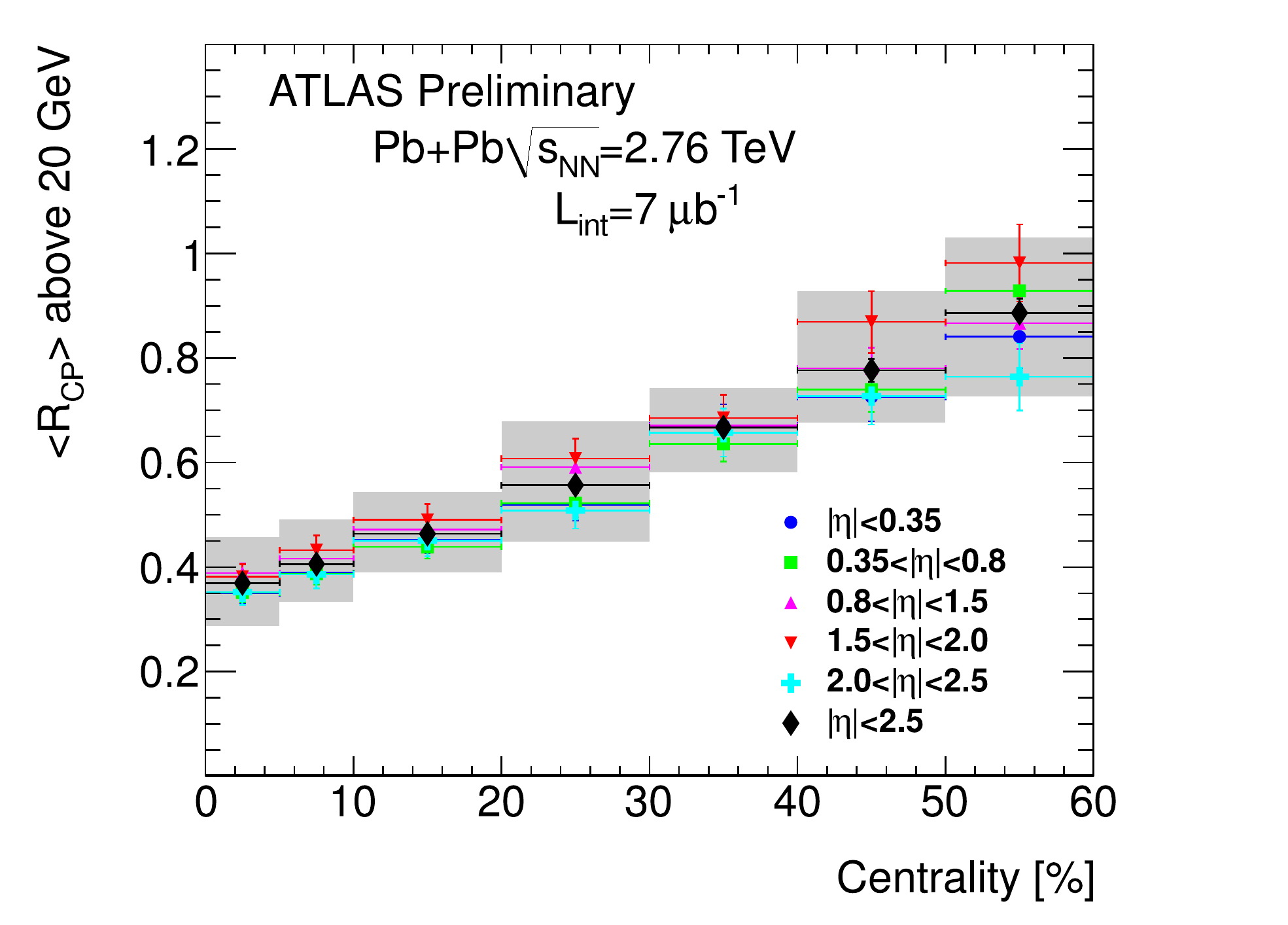}
\caption{\label{charged_rcp_20GeV}
$R_{\mathrm{CP}}$ for $p_T = 20-30$ GeV as a function of centrality, all relative
to the 60-80\% centrality interval.
}
\end{center}
\end{figure}

To study this phenomenon further, charged particle yields have been measured
out to $|\eta| = 2.5$ and $\pT = 30$ GeV~\cite{QM2011spec}.  Tracks are reconstructed in
the full inner detector (ID), with stringent requirements on the number of hits
per track in order to suppress fake tracks.  With these requirements,
efficiencies are typically 70-80\% while fake rates near $\eta=0$ are
less than a percent, although both efficiency and purity are degraded somewhat
going to forward rapidity.  While tracks are measured in ATLAS at transverse momenta
well past 30 GeV, the measured track impact parameter errors (which are also used to reject
fake tracks) are not the same in the data and in the present ATLAS simulations.

The top three panels of Fig.~\ref{charged_rcp} show the corrected spectra out to 30 GeV,
in three inclusive pseudorapidity intervals ($|\eta|<0.35$, $|\eta|<0.8$ and $|\eta|<2.5$).
The statistical errors are shown by the error bars (which are typically smaller than the
graph markers) and total systematic uncertainties are shown by the light grey bands.
Dividing the spectra by the yields measured in the 60-80\% interval as well as the ratios
of the number of binary collisions gives the charged-hadron $R_{CP}$ values shown in
the bottom three panels of Figure~\ref{charged_rcp}, for three centrality intervals
relative to the 60-80\% centrality interval.  The trend for the 0-5\% interval is similar
to that seen in charged particle $R_{AA}$ measured by ALICE~\cite{Aamodt:2010jd}.
There is a peak at 2-3 GeV, a minimum around 7 GeV and then a steady 
rise going out to high $\pT$ which has not yet leveled-off even at $\pT = 30$ GeV.
The value of $R_{CP}$ averaged over $\pT$ from 20 to 30 GeV, shown in Fig.~\ref{charged_rcp_20GeV}, is found to decrease smoothly 
from a value near unity for 50-60\% to approximately 0.4 for the 0-10\% most central collisions.
This trend is both qualititatively and quantitatively similar to the one previously discussed
for fully reconstructed jets.

\section{Conclusions}

Results are presented from the ATLAS detector using over
9 $\mu$b$^{-1}$ of lead-lead collisions from the 2010 LHC heavy ion run.
The charged particle multiplicity
increases by a factor of more than two relative to the top RHIC energy, with a 
centrality dependence very similar to that already measured at RHIC.
Measurements of elliptic flow out to large transverse
momentum also show similar results to what was measured at RHIC.
Measurements of higher harmonics have also been made,
which are able to explain structures in the two-particle correlation
(e.g. the ``ridge'' and ``cone'' phenomena)
previously attributed to jet-medium interactions.
Single muons at high momentum are used to extract the yields of $W^{\pm}$ bosons
as a function of centrality,
which are found to be consistent with binary collision scaling.
Conversely, jets are found to be suppressed in central events by a factor of
two relative to peripheral events, 
with no significant dependence on the jet energy.
Fragmentation functions
are also found to be the same in central and peripheral events.
Updated asymmetry results are presented for two different jet radii and
improved background subtraction, confirming the first ATLAS results.
Finally, charged hadrons have been measured out to 30 GeV, and the measured
values of $R_{CP}$ are found to have a centrality
dependence (relative to peripheral events) similar to that found for jets.

\begin{acknowledgments}
The author would like to thank Derek Teaney and the DPF Heavy Ion session organizers for the 
invitation to speak at DPF 2011, and the ATLAS collaboration for their continued strong support
of the ATLAS heavy ion physics program.
\end{acknowledgments}

\bigskip 

\end{document}